\documentclass[11pt, letterpaper]{article}
\usepackage{amsfonts,amsmath,amsopn,amssymb,amsthm,bbm,latexsym,mathrsfs,pstricks,verbatim,color}
\let\n\noindent

\font\small=cmr8

\font\tenmsy=msbm10
\font\sevenmsy=msbm10 at 7pt
\font\fivemsy=msbm10 at 5pt
\newfam\msyfam % family 11
\textfont\msyfam=\tenmsy
\scriptfont\msyfam=\sevenmsy
\scriptscriptfont\msyfam=\fivemsy

\let\s\sigma
\let\L\Langel

\let\l\left
\let\r\right

\def\z{{\cal Z}}
\let\lf\lfloor
\let\rf\rfloor
\let\lc\lceil
\let\rc\rceil

\def\p{{\tilde p}}

\def\y{{\infty}}

\let\Rw\Rightarrow
\def\l{{\left}}
\def\r{{\right}}

\def\rw{\rightarrow}

\def\L{\langle}

\let\ka\kappa

\font\small=cmr8

\begin{document}

\vskip18pt

\title{\vskip60pt {\bf Fermi-gas interpretation of the  RSOS path representation of the  superconformal unitary minimal  models}}

\vskip18pt

\smallskip
\author{ \bf{P. Jacob$^{1,2}$ and P.
Mathieu$^2$}\thanks{patrick.jacob.1@phy.ulaval.ca,
pmathieu@phy.ulaval.ca.} \\ 
\\
${}^1$Department of Mathematical Sciences, \\University of Durham, Durham, DH1 3LE, UK\\
and\\
${}^2$D\'epartement de physique, de g\'enie physique et d'optique,\\
Universit\'e Laval,
Qu\'ebec, Canada, G1K 7P4.}

\vskip .2in
\bigskip
%\bigskip
%\bigskip
%\bigskip
%\date{August 2007}

\maketitle

\let\Rw\Rightarrow
\let\rw\rightarrow
\let\l\left
\let\r\right
\let\s\sigma
\let\ka\kappa
\let\de\delta

\def\M{{\cal M}}
\def\SM{{\cal SM}}

 \let\g\gamma

\def\lp{{\bar {{\rm P}}^{[k-\frac12]}}}

\def\LP{{ {\rm P}^*}^{[k]}}

% \begin{abstract}

\vskip0.3cm
\centerline{{\bf ABSTRACT}}
\vskip18pt
We derive new  finitized fermionic characters for the  superconformal unitary minimal  models by interpreting the  RSOS configuration sums as  fermi-gas partition functions. This extends to the supersymmetric case the method introduced by Warnaar for the Virasoro unitary mimimal models. The key point in this construction  is the proper  identification of  fermi-type charged particles in terms of the path's peaks. For this, an instrumental preliminary step is  the adaptation to the superconformal case of the operator description of the usual RSOS paths introduced recently.

% for fermionic-type quasi-particles with simple exclusion path representation

%\newpage

%==============================================================================

\section{Introduction}

The solution of the Andrews-Baxter-Forrester and the Forrester-Baxter restricted-solid-on-solid (RSOS) models \cite{ABF,FB} by the corner-transfer matrix method leads to an expression for the 
 local state probability of the order variable in terms of a configuration sum. Each such configuration sum provides thus a finitization of the character of an irreducible module of the corresponding minimal model.\footnote{These statistical  models are related to the minimal models \cite{Huse,Rig,Nak}. More precisely,  in their scaling limit, the statistical models correspond to conformal field theories only at criticality. The minimal models are associated to the transition from regime III to IV.  But even off-criticality, the configuration sums describe conformal characters \cite{Kyotoa,Kyoto, Krev}. This remarkable off-critical relationship is explained in \cite{SB,SN}. In this way, the characters of the minimal models $\M(p',p)$ are related to the configuration sums  in regime III.}   In this description, every state is represented by a particular configuration   \cite{Kyoto,Mel}. Each configuration is naturally interpreted as a lattice path \cite{OleJS,FLPW}. The finitization parameter is the path length $L$. The usual Virasoro characters are recovered in the limit $LÊ\rw \y$.

The path description is combinatorial:  counting the paths gives directly the character without subtraction. This provides thus a royal road for the derivation of fermionic expressions (cf.  \cite{KKMMa,KKMMb,Mel}) of the characters. 
%as lead to a fruitful description of the

Various lines of attack for deriving fermionic formulae from configuration sums have been proposed.
In a seminal contribution, Melzer \cite{Mel} has conjectured many finitized fermionic expressions for the unitary minimal models (extending significantly  the original conjectures of \cite{KKMMb}).  The first few conjectured positive multiple-sum formulae were proved by demonstrating that  they satisfy the same recurrence relations that characterize the corresponding configuration sums. This method of proof has been generalized  and made into a powerful technique (under the name of telescopic expansion method) in \cite{Ber}, where many fermionic formulae (those for modules $(r,s)$ with $s=1$) for all Virasoro unitary models are demonstrated. (All cases are covered in \cite{Sch}.) Note however that this telescopic expansion method is essentially a verification tool  that requires a candidate expression for the fermionic form.  But in \cite{Ber}, the fermionic expressions are presented as  natural $q$-deformations of   state counting problems for a specific (RSOS motivated) truncation of the space of states of the thermodynamic limit of the quantum XXZ spin chain \cite{BR}. This observation has proved to be fruitful. In \cite{BMlmp}  many new fermionic formulae (i.e., for all minimal models) are conjectured, all  motivated by this counting procedure. These were successively generalized and  proved in \cite{BMS} by verifying that they satisfy RSOS-type recurrence relations. Note however that these expressions are related to paths in a very indirect way.
%urence relations which are proved from polynomial versions of the characters  but those are extracted from a  truncation of the space of states of the quantum XXZ spin chain instead of the RSOS model, and in that sense, they do not originate from paths.} 

A  frontal attack of the difficult combinatorics of the paths for the generic Forrester-Baxter models is considered in \cite{FWa,FLPW,FW,Wel} (albeit using variables that are characteristics of the XXZ spectrum analysis and whose path interpretation is not immediate). The key preliminary step is the discovery of a new (manifestly positive definite) characterization of the weight  of a  Forrester-Baxter RSOS path (cf. App. A of \cite{FLPW}). From then on, the  strategy followed in these works is to describe a generic path pertaining to the (finitized version of the) minimal model $\M(p', p)$ in terms of successive transformations acting on the unique and trivial path for the (formal) $\M(1,3)$ model. This relies on two  explicit combinatorial transformations defined directly on the paths: a Bressoud-type transformation \cite{BreL} that relates paths within a family defined by $\M(p',p+kp')$ for different values of $k\geq 1$, and a duality transformation \cite{BMlmp} that  relates $\M(p',p)$ to $\M(p-p',p)$.  Although these  basic transformations are quite intuitive, the resulting construction turns out to be technically rather involved. Nevertheless, all characters for all the minimal models have been written in fermionic form along this line \cite{Wel}.  

For the unitary models, the combinatorics of the RSOS paths is considerably simplified. In that case,  Warnaar has shown that a configuration sum can be regarded as a (grand-canonical) partition function of a one-dimensional gas of fermi-type particles subject to restriction rules \cite{OleJS,OleJSb}. This method leads to the fermionic expression of the characters in a  simple, direct and totally constructive way. Moreover, in this simpler context, the procedure can be formulated in terms of variables that have a clear path interpretation.\footnote{For completeness, it should be added that a fermi-gas description has been obtained also for the $\M(2,p)$  models in \cite{W97}, but not directly from RSOS paths. Moreover, we have shown recently that the non-unitary minimal models of the type $\M(k+1,2k+3)$ do  have a path representation  similar to that of the unitary minimal models and thus an analogous fermi-gas-type representation \cite{JSTAT}. However, these specific paths do not (yet) originate from a RSOS model and their relation to the known RSOS paths (i.e, by means of a bijection) has not been established so far.}
% description that has not been made equivalent to the RSOS one.}
 %However, it has a severe drawback in that it applies only to the unitary models.

Toward the goal of unraveling the conformal-field-theoretical quasi-particle formalism underlying  the fermionic formulae, this fermi-gas method  seems to be promising: The charged particles identified within the path would seem  to be the very quasi-particles that should be lifted to the operator level. A step in this direction 
%Further support for this expectation 
is presented in \cite{JMop}, where an operator interpretation of these paths is presented.

In contrast to the Virasoro minimal models, the superconformal ones have been little studied from the path perspective. In fact, the RSOS models underlying the non-unitary cases have  not yet been formulated. The unitary case, however, is a special case of a general class of RSOS models solved in \cite{Kyoto}. The corresponding finitized fermionic formulae for the unitary superconformal models have been conjectured in \cite{BG} by finitizing the expression of the conjectured vacuum module character of \cite{KKMMb} and extending it to all modules.
% for the vacuum module) as the natural finitized version of the on the basis of the RSOS path representation.
These expressions have subsequently been proved in \cite{Sch} using the telescopic expansion method.

The aim of the present article is to work out the combinatorics
of  the superconformal minimal models from their RSOS path representation, by extending directly the fermi-gas approach of \cite{OleJS}.
The conditions defining the allowed RSOS configurations (and, in consequence, the corresponding paths) are reviewed in section 2. The very crucial step in this approach is to formulate a precise and non-ambiguous method for  interpreting the path as a mixture of charged fermi-type particles.  This requires a criterion, first, for attributing a charge to an isolated particle and then for identifying its various particle components within a charge complex. This is not quite straightforward in the superconformal context given that the path is typically composed not only of North-East  and South-East  edges (the only type of edges allowed in their non-supersymmetrical counterpart), but also horizontal  ones.

 Our first approach to the problem was  in continuity with our recent operator construction of the paths representing the states in the unitary Virasoro minimal models \cite{JMop}. This method, when abstracted from its path  scaffolding, does not lead naturally to finitized characters but rather to their infinite length limit, i.e., the Virasoro characters. We have succeeded in extending this approach to the superconformal case. By trying to understand the finitized version of the resulting expressions, we were led to the particle identification proposed here. This is taken as our starting point in this article. (The operator construction is presented in section 4.) Once the particle interpretation of a path is understood, the rest of the analysis follows essentially the  main steps of the construction  presented in  \cite{OleJS}.
 
  The resulting fermionic formulae appear to be new in their finitized version. However, after a simple transformation,  they reduce to the known expressions in the infinite length limit \cite{BG,Sch}.  But our main point is not much that novel finitized forms are generated. It is rather  that the paths  are now dressed with a  clear particle interpretation. Consequently  these fermionic formulae are obtained by a fully constructive method.
 
 %fermionic characters.

\section{$\SM(k+2,k+4)$ RSOS paths}

A configuration pertaining to the (regime-III) RSOS realization of the finitized $\SM(k+2,k+4)$ unitary minimal models is described by a sequence of values of the height variables $\ell_i\in \{1,2,\cdots, k+3\}$. The height index  is bounded by  $0\leq i\leq L$. Adjacent heights are subject to the admissibility condition:
\begin{align}\label{adm}
&\ell_i-\ell_{i+1}\in \{-2,0,2\},\nonumber\\
&\ell_i+\ell_{i+1}\in \{4,6,\cdots , 2(k+2)\}.
\end{align}
Each configuration is specified by particular boundary conditions: the values of $\ell_0$ and $\ell_L$. A configuration  is weighted by the expression
\begin{equation}\label{wei}
w=\sum_{i=1}^{L-1} \frac{i}4 |\ell_{i-1}-\ell_{i+1}| .
\end{equation}

A path is the contour of a configuration. It is thus a sequence of edges joining the adjacent vertices $(i,\ell_i)$ and $(i+1,\ell_{i+1})$ of the configuration. These edges can be either  North-East (NE), South-East (SE) or  horizontal (H), corresponding to the cases where $\ell_{i+1}-\ell_i=2,\, -2$ or $0$ respectively. Note however, that an H edge linking two heights both equal to 1 or both equal to $k+3$ is not allowed: H edges on the boundaries of the rectangular strip delimitating the paths are forbidden. 

The expression (\ref{wei}) for  the weight of a configuration applies directly  to the corresponding path. It follows that vertices at extremal positions in the paths (either minima or maxima) do not contribute. Vertices at position $i$  in-between two NE or two SE edges contribute to $i$, while those in-between an H edge and a non-H edge (in both orders) contribute to $i/2$ (cf. Fig. {\ref{fig1}). 
%(Figure des differentes sequences de 2 liens et de leur poids.)

%%%%%%%%%%%%%%%%%%%%%%
%%%% Figure 1 Fermionic Gaz Susy                          %%%%
%%%%%%%%%%%%%%%%%%%%%%%%%%%%%%%%%%%%%%%%%%%%%%%%%%%%%%%%%%%%%%

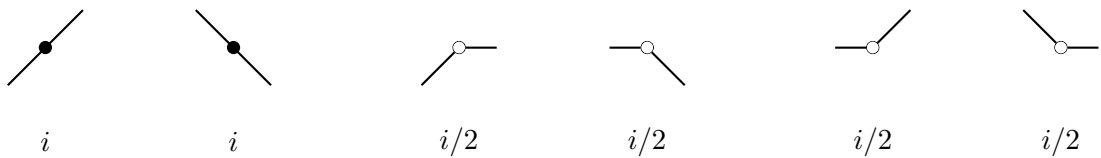
\begin{figure}[ht]
\caption{{\footnotesize The four types of contributing vertices and their weight. In all cases, it is understood that the vertex horizontal position is $i$. Here and in some of the following figures, the two type of vertices are denoted by black dots (with weight $i$) or circles (with weight  $i/2$). The other  vertices, namely those corresponding to local path extrema, have zero weight.}} \label{fig1}
\begin{center}
\begin{pspicture}(0,0)(16.0,2.0)
%GRAPHIC
%first part
\psline{-}(0.5,0.5)(1.0,1.0) \psline{-}(1.0,1.0)(1.5,1.5)
\psline{-}(3.0,1.5)(3.5,1.0) \psline{-}(3.5,1.0)(4.0,0.5)
%second part
\psline{-}(6.0,0.5)(6.5,1.0) \psline{-}(6.5,1.0)(7.0,1.0)
\psline{-}(8.5,1.0)(9.0,1.0) \psline{-}(9.0,1.0)(9.5,0.5)
%third part
\psline{-}(11.5,1.0)(12.0,1.0) \psline{-}(12.0,1.0)(12.5,1.5)
\psline{-}(14.0,1.5)(14.5,1.0) \psline{-}(14.5,1.0)(15.0,1.0)
%arrows
%\psline{->}(5.5,1.5)(6.0,1.5) \psline{->}(11.0,1.5)(11.5,1.5)
%dots
\psset{dotsize=5pt}\psset{dotstyle=*}
\psdots(1.0,1.0)(3.5,1.0)
\psset{dotsize=5pt}\psset{dotstyle=o}
\psdots(6.5,1.0)(9.0,1.0)(12.0,1.0)(14.5,1.0)
%text
\rput(1.0,-0.25){{\small $i$}} \rput(3.5,-0.25){{\small $i$}}
\rput(6.5,-0.25){{\small $i/2$}} \rput(9.0,-0.25){{\small $i/2$}}
\rput(12.0,-0.25){{\small $i/2$}} \rput(14.5,-0.25){{\small $i/2$}}
\end{pspicture}
\end{center}
\end{figure}

The weighted sum over all paths with specified boundaries is the configuration sum
\begin{equation}
X_{\ell_0, \ell_L}(q) = \sum_{\substack{\text{paths with fixed end}\\\text{points $\ell_0$ and $\ell_L$}}}q^w.
%~($  \ell_0$ $\ell_L$ fixed}}
%\sum_{i=1}^{L-1} q^{\frac14 |\ell_{i-1}-\ell_{i+1}|} \; ,
\end{equation}
% \ell_0\sim\ell_1\\ \ell_{L-1}\sim\ell_L} 
% where $\ell_i\sim\ell_{i+1}$ means that $\ell_i$ and $\ell_{i+1}$ satisfy the two admissibility conditions (\ref{adm}).
With the proper relation between $(\ell_0,\ell_L)$ and the irreducible indices $(r,s)$, where $1\leq r\leq k+1$ and $1\leq s\leq k+3$, 
%(with modules labeled by $(r,s)$ and $(k+2-r,k+4-s)$ being identified), 
%
this is the finitized version of the superconformal characters.

%%%%%%%%%%%%%%%%%%%%%%%%%%%%%%%%%%%%%%%%%%%%%%%%%%%%%%%%%%%%%%
%%%% Figure 2 Fermionic Gaz Susy                          %%%%
%%%%%%%%%%%%%%%%%%%%%%%%%%%%%%%%%%%%%%%%%%%%%%%%%%%%%%%%%%%%%%

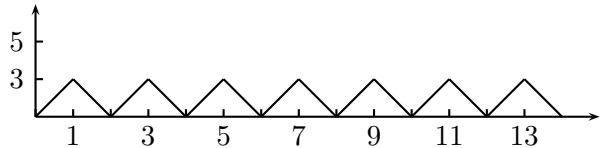
\begin{figure}[ht]
\caption{{\footnotesize The path of lowest energy  for all finitized superconformal unitary models $\SM(k+2,k+4)$. This path is associated to the vacuum state. Note that here and in all other figures,  the vertical axis has been rescaled by a factor 2. The height of a peak refers to this rescaled height. The path is thus a sequence of peaks of height 1 whose number $n_1$ is fixed by the length $L$, here equal to 12, via $L=2n_1$.}}
\label{fig2}
\begin{center}
\begin{pspicture}(0,0)(9.5,2.5)
%axis
\psline{->}(0.5,0.5)(0.5,2.0) \psline{->}(0.5,0.5)(8.0,0.5)
%\psset{linestyle=dashed,dashadjust=false} \psline(0.5,2.0)(15.5,2.0)
%\psset{linestyle=dotted}
%\psset{linestyle=solid}
%\psline{<->}(4.5,1.0)(6.5,1.0)
\psset{linestyle=solid}
%units
\psline{-}(0.5,0.5)(0.5,0.6) \psline{-}(1.0,0.5)(1.0,0.6)
\psline{-}(1.5,0.5)(1.5,0.6) \psline{-}(2.0,0.5)(2.0,0.6)
\psline{-}(2.5,0.5)(2.5,0.6) \psline{-}(3.0,0.5)(3.0,0.6)
\psline{-}(3.5,0.5)(3.5,0.6) \psline{-}(4.0,0.5)(4.0,0.6)
\psline{-}(4.5,0.5)(4.5,0.6) \psline{-}(5.0,0.5)(5.0,0.6)
\psline{-}(5.5,0.5)(5.5,0.6) \psline{-}(6.0,0.5)(6.0,0.6) 
\psline{-}(6.5,0.5)(6.5,0.6) \psline{-}(7.0,0.5)(7.0,0.6) 
\rput(1.0,0.25){{\small $1$}} \rput(2.0,0.25){{\small $3$}}
\rput(3.0,0.25){{\small $5$}} \rput(4.0,0.25){{\small $7$}}
\rput(5.0,0.25){{\small $9$}} \rput(6.0,0.25){{\small $11$}}
\rput(7.0,0.25){{\small $13$}}
 \psline{-}(0.5,1.0)(0.6,1.0)
\psline{-}(0.5,1.5)(0.6,1.5) 

\rput(0.25,1.0){{\small $3$}} \rput(0.25,1.5){{\small $5$}}
%graphic
\psline{-}(0.5,0.5)(1.0,1.0) \psline{-}(1.0,1.0)(1.5,0.5)
\psline{-}(1.5,0.5)(2.0,1.0) \psline{-}(2.0,1.0)(2.5,0.5)
\psline{-}(2.5,0.5)(3.0,1.0) \psline{-}(3.0,1.0)(3.5,0.5)
\psline{-}(3.5,0.5)(4.0,1.0) \psline{-}(4.0,1.0)(4.5,0.5)
\psline{-}(4.5,0.5)(5.0,1.0) \psline{-}(5.0,1.0)(5.5,0.5)
\psline{-}(5.5,0.5)(6.0,1.0) \psline{-}(6.0,1.0)(6.5,0.5)
\psline{-}(6.5,0.5)(7.0,1.0) \psline{-}(7.0,1.0)(7.5,0.5)

%dots
%\psset{dotsize=3pt}
%\psdots(2.5,1.0)(3.0,1.5)(4.0,1.5)(5.0,1.5)(6.0,1.5)(6.5,1.0)(10.0,1.0)(11.0,1.0)

\end{pspicture}
\end{center}
\end{figure}

To a large extend, the combinatorics of the paths is independent of the boundary conditions. To avoid unnecessary complications, we will thus confine ourself  to the  analysis of the simple case where $\ell_0=\ell_L=1$. These conditions characterize the vacuum module. With the path extremities fixed in this way,  the only allowed heights are the odd numbers between 1 and $k+3$. It is thus convenient to reduce the vertical scale by a factor of two as shown in Fig. \ref{fig2} where the path describing the vacuum state (for all unitary minimal models)  is displayed. Its  weight is readily seen to be zero. 
%(C'est la sequence de pics 1). 
A generic path for a model with $k>3$ is presented in Fig. \ref{fig3}.

%%%%%%%%%%%%%%%%%%%%%%%%%%%%%%%%%%%%%%%%%%%%%%%%%%%%%%%%%%%%%%
%%%% Figure 3 Fermionic Gaz Susy                          %%%%
%%%%%%%%%%%%%%%%%%%%%%%%%%%%%%%%%%%%%%%%%%%%%%%%%%%%%%%%%%%%%%

\begin{figure}[ht]
\caption{{\footnotesize A generic path valid for all models with $k \geq$ 4. The procedure for determining its particle content is explained is section 3. There are three complexes, delimited by the four points on the horizontal axis.  The path can be decomposed into a sequence of charged particles. The charge of the highest peak of a complex (or the leftmost one if it is not unique) is its height plus 1/2 if it is topped by at least one horizontal (H) edge. For  instance, the peak centered on the horizontal position 7/2 has charge 5/2. Note that extra H edges at its top would not modify its charge. Similarly,  the one at position 17 has charge 3 and that at 28 has charge 1. For the others,  as explained in section 3.2, the charge is read from the top to the indicated baseline (drawn as a dotted line a bit below its actual position for clarity). This charge assignment has to be adjusted if the particle incorporates an H edge at the height of the baseline (cf. section 3), in which case this increases its charge by 1/2. The arrows of the dotted lines delimitated the actual particles; the length of these dotted lines is the particle diameter, which is also twice the particle charge. The remaining H edges are particles of charge 1/2.  (In order to avoid any ambiguities with our subsequent analysis, note that in the following, we will forbid the insertion of charge 1/2 particles in-between two particles. In the present context, it means that it is the H edge between $i=5$ and 6 that belongs to the charge 3/2 particle and the subsequent H edge represents a charge 1/2 particle inserted within this charge 3/2 particle. Of course, this reinterpretation does not modify the charge content.)  If $n_j$ stands for the number of particles of charge $j$, the particle content of this path is $n_{\frac12} = n_{\frac32}= 3, n_1=n_2=n_{\frac52}=n_3=1$. The length is $L=\sum jn_j = 29$.}}
\label{fig3}
\vskip0,3cm
\begin{center}
\begin{pspicture}(0,0)(15.5,3.6)
%axis
\psline{->}(0.5,0.5)(0.5,3.5) \psline{->}(0.5,0.5)(15.5,0.5)
%\psset{linestyle=dashed,dashadjust=false} \psline(0.5,2.0)(15.5,2.0)
%\psset{linestyle=dotted}
%\psset{linestyle=solid}
%\psline{<->}(4.5,1.0)(6.5,1.0)
\psset{linestyle=solid}
%units
\psline{-}(0.5,0.5)(0.5,0.6) \psline{-}(1.0,0.5)(1.0,0.6)
\psline{-}(1.5,0.5)(1.5,0.6) \psline{-}(2.0,0.5)(2.0,0.6)
\psline{-}(2.5,0.5)(2.5,0.6) \psline{-}(3.0,0.5)(3.0,0.6)
\psline{-}(3.5,0.5)(3.5,0.6) \psline{-}(4.0,0.5)(4.0,0.6)
\psline{-}(4.5,0.5)(4.5,0.6) \psline{-}(5.0,0.5)(5.0,0.6)
\psline{-}(5.5,0.5)(5.5,0.6) \psline{-}(6.0,0.5)(6.0,0.6)
\psline{-}(6.5,0.5)(6.5,0.6) \psline{-}(7.0,0.5)(7.0,0.6)
\psline{-}(7.5,0.5)(7.5,0.6) \psline{-}(8.0,0.5)(8.0,0.6)
\psline{-}(8.5,0.5)(8.5,0.6) \psline{-}(9.0,0.5)(9.0,0.6)
\psline{-}(9.5,0.5)(9.5,0.6) \psline{-}(10.0,0.5)(10.0,0.6)
\psline{-}(10.5,0.5)(10.5,0.6) \psline{-}(11.0,0.5)(11.0,0.6)
\psline{-}(11.5,0.5)(11.5,0.6) \psline{-}(12.0,0.5)(12.0,0.6)
\psline{-}(12.5,0.5)(12.5,0.6) \psline{-}(13.0,0.5)(13.0,0.6)
\psline{-}(13.5,0.5)(13.5,0.6) \psline{-}(14.0,0.5)(14.0,0.6)
\psline{-}(14.0,0.5)(14.0,0.6) \psline{-}(14.5,0.5)(14.5,0.6)
%\psline{-}(15.0,0.5)(15.0,0.6) \psline{-}(15.5,0.5)(15.5,0.6)
\rput(3.0,0.25){{\small $5$}} 
\rput(5.5,0.25){{\small $10$}} \rput(8.0,0.25){{\small $15$}}
\rput(10.5,0.25){{\small $20$}} \rput(13.0,0.25){{\small $25$}}
 \psline{-}(0.5,1.0)(0.6,1.0)
\psline{-}(0.5,1.5)(0.6,1.5) \psline{-}(0.5,2.0)(0.6,2.0)
\psline{-}(0.5,2.5)(0.6,2.5) \psline{-}(0.5,3.0)(0.6,3.0)
\rput(0.25,1.0){{\small $3$}} \rput(0.25,2.0){{\small $7$}}
 \rput(0.25,3.0){{\small $11$}}
%graphic
\psline{-}(0.5,0.5)(1.0,1.0) \psline{-}(1.0,1.0)(1.5,1.0)
\psline{-}(1.5,1.0)(2.0,1.5) \psline{-}(2.0,1.5)(2.5,1.5)
\psline{-}(2.5,1.5)(3.0,1.0) \psline{-}(3.0,1.0)(3.5,1.0)
\psline{-}(3.5,1.0)(4.0,1.0) \psline{-}(4.0,1.0)(4.5,1.5)
\psline{-}(4.5,1.5)(5.0,1.0) \psline{-}(5.0,1.0)(5.5,0.5)
\psline{-}(5.5,0.5)(6.0,1.0) \psline{-}(6.0,1.0)(6.5,1.5)
\psline{-}(6.5,1.5)(7.0,1.5) \psline{-}(7.0,1.5)(7.5,1.0)
\psline{-}(7.5,1.0)(8.0,1.0) \psline{-}(8.0,1.0)(8.5,1.5)
\psline{-}(8.5,1.5)(9.0,2.0) \psline{-}(9.0,2.0)(9.5,1.5)
\psline{-}(9.5,1.5)(10.0,1.5)
 \psline{-}(10.0,1.5)(10.5,2.0)\psline{-}(10.5,2.0)(11.0,1.5)
\psline{-}(11.0,1.5)(11.5,1.5) \psline{-}(11.5,1.5)(12.0,1.0)
\psline{-}(12.0,1.0)(12.5,1.0) \psline{-}(12.5,1.0)(13.0,1.5)
\psline{-}(13.0,1.5)(13.5,1.0) \psline{-}(13.5,1.0)(14.0,0.5)
\psline{-}(14.0,0.5)(14.5,1.0) \psline{-}(14.5,1.0)(15.0,0.5)

%particles
\psset{linestyle=dotted}
 
 \psline{<-}(3.5,1.0)(3.7,0.9)\psline{-}(3.7,0.9)(4.8,0.9)\psline{->}(4.8,0.9)(5.0,1.0)
  \psline{<-}(6.0,1.0)(6.2,0.9)\psline{-}(6.2,0.9)(7.8,0.9)\psline{->}(7.8,0.9)(8.0,1.0)
   \psline{<-}(9.5,1.5)(9.7,1.4)\psline{-}(9.7,1.4)(10.8,1.4)\psline{->}(10.8,1.4)(11.0,1.5)
  \psline{<-}(12.0,1.0)(12.2,0.9)\psline{-}(12.2,0.9)(13.3,0.9)\psline{->}(13.3,0.9)(13.5,1.0)

%dots
%\psset{dotsize=3pt}
%\psdots(2.5,1.0)(3.0,1.5)(4.0,1.5)(5.0,1.5)(6.0,1.5)(6.5,1.0)(10.0,1.0)(11.0,1.0)

\end{pspicture}
\end{center}
\end{figure}

\section{The combinatorics of the $\SM(k+2,k+4)$ RSOS paths}

\subsection{ The path as a sequence of charged peaks}

The starting point of the analysis is the interpretation of a path as being filled by  a particular distribution of one-dimensional interacting charged 
particles. The particles are the `basic peaks'  and their charge is related to their height, with the height measured in terms of the reduced vertical scale. 

The proper charge characterization of  an isolated peak, namely, a peak delimitated by two points on the horizontal axis, is the following.  A peak of charge $j$ has diameter $2j$ and height $\lfloor j\rfloor$.\footnote{As usual, $\lfloor j\rfloor$ denotes the largest integer smaller than $j$.} The charge  $j$ can take both integer and half-integer values and it is bounded by $1\leq j\leq k/2+1$. 

A particle with integer charge $j$  corresponds  to a peak described by a triangle with $j $ NE and $j$ SE edges.  A particle of half-integer charge is described by a  flatten triangle  with  $\lfloor j\rfloor$ NE edges followed by one H edge and  $\lfloor j\rfloor$ SE edges;  the triangle is thus topped by one (and only one) H  edge. Clearly, the introduction of half-integer charges is forced by the presence of H links in the paths. The simplest examples are drawn in Fig. \ref{fig4}.

%%%%%%%%%%%%%%%%%%%%%%%%%%%%%%%%%%%%%%%
%%%% Figure 4 Fermionic Gaz Susy                          %%%%
%%%%%%%%%%%%%%%%%%%%%%%%%%%%%%%%%%%%%%%%%%%%%%%%%%%%%%%%%%%%%%

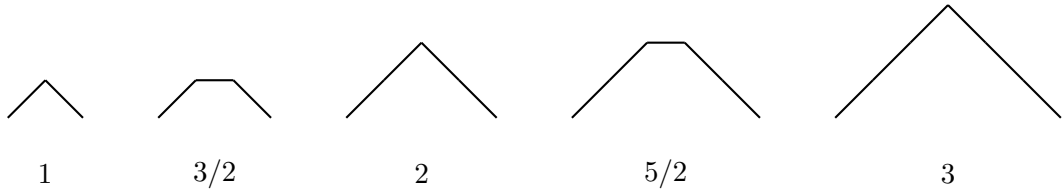
\begin{figure}[ht]
\caption{{\footnotesize The first few particle shapes together with their charge assignment. }} \label{fig4}
\begin{center}
\begin{pspicture}(0,0)(16.0,2.0)

%GRAPHIC
%first part
\psline{-}(0.5,0.5)(1.0,1.0) \psline{-}(1.0,1.0)(1.5,0.5)

\psline{-}(2.5,0.5)(3.0,1.0) \psline{-}(3.0,1.0)(3.5,1.0)
\psline{-}(3.5,1.0)(4.0,0.5) 

%second part
\psline{-}(5.0,0.5)(5.5,1.0) \psline{-}(5.5,1.0)(6.0,1.5)
\psline{-}(6.0,1.5)(6.5,1.0) \psline{-}(6.5,1.0)(7.0,0.5)

\psline{-}(8.0,0.5)(8.5,1.0) \psline{-}(8.5,1.0)(9.0,1.5)
\psline{-}(9.0,1.5)(9.5,1.5) \psline{-}(9.5,1.5)(10.0,1.0)
\psline{-}(10.0,1.0)(10.5,0.5)

%third part
\psline{-}(11.5,0.5)(12.0,1.0) \psline{-}(12.0,1.0)(12.5,1.5)
\psline{-}(12.5,1.5)(13.0,2.0) \psline{-}(13.0,2.0)(13.5,1.5)
\psline{-}(13.5,1.5)(14.0,1.0) \psline{-}(14.0,1.0)(14.5,0.5)

%arrows
%\psline{->}(5.5,1.5)(6.0,1.5) \psline{->}(11.0,1.5)(11.5,1.5)

%dots
%\psset{dotsize=5pt}\psset{dotstyle=*}
%\psdots(1.0,1.0)(3.5,1.0)
%\psset{dotsize=5pt}\psset{dotstyle=o}
%\psdots(6.5,1.0)(9.0,1.0)(12.0,1.0)(14.5,1.0)

%text
\rput(1.0,-0.25){{\small $1$}} \rput(3.25,-0.25){{\small $3/2$}}
\rput(6.0,-0.25){{\small $2$}} \rput(9.25,-0.25){{\small $5/2$}}
\rput(13.0,-0.25){{\small $3$}} 

\end{pspicture}
\end{center}
\end{figure}

The rationale underlying this characterization of the basic particles is the following. As the parameter $k$ is   increased, one expects the number of basic particles to increase. (For the following discussion, it should always be kept in mind  that the vertical axis has been  rescaled by 2, that is, the height between 1 and 3 is  rescaled to 1).  For $k=0$, only triangles of height 1 are allowed. This is a trivial model with a single state represented by the path zig-zagging between 1 and 3, the state illustrated in Fig. \ref{fig2}. The maximal height is also 1 when $k=1$ except that now an H edge is permitted at height 1. It is thus natural to associate to this new allowed particle a charge differing from that of the sole particle appearing in the $k=0$ case. Therefore,   the  flatten version of the triangle of height 1 is attributed charge 3/2.
As $k$  is changed from 1 to 2, there is another possibility, which is a particle of charge 2 corresponding to a triangle of height 2. However its flatten version is not allowed since an H edge at the top of the path's  bordering  strip  is ruled out by the condition (\ref{adm}). As $k$ is changed from 2 to 3,  a peak of height 3 is not allowed but the flatten version of the peak of height 2 is now possible; it is given charge 5/2. We thus see that as $k$ increases by one, there is a new allowed charge for the particles: to the charges $1,\cdots, k/2+1$, we add $(k+1)/2+1$. 

Summing up, the charge $j$ of  an isolated particle  is  related to its height $h$ and the number (0 or 1) of its top H edge as $j=h+(\# H)/2$. The allowed values of these charges are $1,\cdots, k/2+1$.
%Equivalently, we could count an isolated top H edge as a height 1/2

In addition to the particles of charge $\geq 1$, we introduce  particles of charge 1/2. These, obviously, cannot be isolated by two points on the horizontal axis as the height $h$ of an H edge must be strictly greater than 1. Moreover, these particles must be distinguished from the top H edge of the particles of half-integer charge. In other words, a particle of charge $5/2$ say, is not a composite of  particles of charge 2 and 1/2.

\subsection{Particle interactions}

As in standard RSOS paths \cite{OleJS}, identical particles have a hard-core repulsion. The smallest distance between two particles with  the same integer  charge is equal to the diameter of one particle, i.e., $2j$ if they have charge $j$.  The distance is measured from center to center, that is, from one peak to another.
This also  holds  for half-integer charges if the distance is measured from the middle of the particle, namely, the position of the middle of the top H edge. For instance, the separation between two 5/2 particles is 5.

In contrast, particles of different charges can interpenetrate.  This operation must preserve the particle identities.
%, that is, the total particle content. 
For the penetration of a particle of charge $i$ within one of larger charge $j$, this means that no height larger  than $j$ should result,  taking  the particles to be isolated.\footnote{That the height of the path does not become larger than  that of the highest peak ensures that this penetration process is well-defined within a model specified by a given value of $k$, which sets the upper bound on the allowed peaks height.} This  implies in turn that there must exist a minimal distance between the 
two particles (measured from their respective center). This minimal distance  is $2i$ if the particle $i$ is at  the right of the largest particle and $2i+1/2$ if it is at its left. The asymmetry comes from the fact that identical configurations are not regarded as distinct: when two peaks have the same height (counting the possible H edge as contributing 1/2 to the height), it is the leftmost peak that is taken to be the highest charge particle. (The general rule for reading the charge content of a path is given below.)

The evaluation of the minimal distance between particles, or more generally, the interaction process, relies on a novel mechanism  of interpenetration (compared to \cite{OleJS}). The novelty, induced by the presence of H edges,  is that the particle with smallest charge can be slightly deformed. 
When the small particle  penetrates the large one from the right (left), it can start (finish) with an H edge, 
whose contribution (1/2) needs to be added to the charge of the deformed peak. If the deformed particle has half-integer charge, it is as if its top H edge was moved  to the beginning (end) of the particle. If the deformed particle has integer charge, because it has an H edge at its beginning (end), the peak height is reduced by 1/2: it has thus a top H edge.
In every cases, it is the peak position or the middle of the top H edge that determines the center of the deformed particle, from which distances are measured.

%can be moved to the beginning (end) of the particle. Similarly, when a particle of integer charge penetrates a larger charge particle from the right (left), its NE-SE top  two edges are changed into two H edges, one of which remains at the top -- whose middle defines the particle center --  and the other is moved to the beginning (end) of the particle. 
% Similarly, when a particle of integer charge penetrates a larger charge particle from the right (left), its NE-SE top  two edges are changed into two H edges, one of which remains at the top -- whose middle defines the particle center --  and the other is moved to the beginning (end) of the particle. In every cases, it is the peak position or the middle of the H edge that determines the center of the particle, from which measures are taken. 

In the various steps of interpenetration, the configuration of the smallest particle alternates between its original form and its deformed version, as exemplified in Fig. \ref{fig5}.
Note that the deformation is not a continuous process: there cannot be more than one H edge at the height of insertion.
This mechanism of successive shape deformation and restoration ensures that the consecutive steps of interpenetration induce a weight difference equal to 1 and a separation difference of 1/2 between the particles. The various cases are illustrated in Fig. \ref{fig5}. 
%(Faire tous les cas de penetration a gauche et a droite de ($i,j)= (2,3),(2,5/2),(3/2,3),(3/2,5/2)$ et ecrire la distance dans chaque cas pour illustrer que ca varie de 1/2.)
Note that a particle of charge 1 cannot be deformed in that way, its number of constituent edges being too small.

%%%%%%%%%%%%%%%%%%%%%%%
%%%%%%%%%%%%%%%%%%%%%%%%%%%%%%%%%%%%%%%%%%%%%%%%%%%%%%%%%%%%%%
%%%% Figure 5 Fermionic Gaz Susy                          %%%%
%%%%%%%%%%%%%%%%%%%%%%%%%%%%%%%%%%%%%%%%%%%%%%%%%%%%%%%%%%%%%%

\begin{figure}[ht]
\caption{{\footnotesize The various stages of interpenetration of a particle of type $i$ through a particle of type $j$ (with $j>i$) for the cases where (a) $(i,j)=(2,3)$, (b) $(i,j)=(2,5/2),$ (c) $(i,j)=(3/2,3)$ and (d) $(i,j)=(3/2,5/2)$. In all cases, the even (second, fourth, etc.) steps display the deformation pattern of the inserted particle. The weight difference between  any configuration and its subsequent one is exactly 1. }} \label{fig5}
\begin{center}
\begin{pspicture}(0,0)(13.0,11.0)

%GRAPHIC 9a1
\psline{-}(0.3,8.8)(0.6,9.1) \psline{-}(0.6,9.1)(0.9,9.4)
\psline{-}(0.9,9.4)(1.2,9.7) \psline{-}(1.2,9.7)(1.5,9.4)
\psline{-}(1.5,9.4)(1.8,9.1) \psline{-}(1.8,9.1)(2.1,8.8)
\psline{-}(2.1,8.8)(2.4,9.1) \psline{-}(2.4,9.1)(2.7,9.4)
\psline{-}(2.7,9.4)(3.0,9.1) \psline{-}(3.0,9.1)(3.3,8.8)

%dots
\psset{dotsize=3pt}\psset{dotstyle=*}
\psdots(0.6,9.1)(0.9,9.4)(1.5,9.4)(1.8,9.1)(2.4,9.1)(3.0,9.1)

%GRAPHIC 9a2
\psline{-}(3.9,8.8)(4.2,9.1) \psline{-}(4.2,9.1)(4.5,9.4)
\psline{-}(4.5,9.4)(4.8,9.7) \psline{-}(4.8,9.7)(5.1,9.4)
\psline{-}(5.1,9.4)(5.4,9.1) \psline{-}(5.4,9.1)(5.7,9.1)
\psline{-}(5.7,9.1)(6.0,9.4) \psline{-}(6.0,9.4)(6.3,9.4)
\psline{-}(6.3,9.4)(6.6,9.1) \psline{-}(6.6,9.1)(6.9,8.8)

%dots
\psset{dotsize=3pt}\psset{dotstyle=*}
\psdots(4.2,9.1)(4.5,9.4)(5.1,9.4)(6.6,9.1)

\psset{dotsize=3pt}\psset{dotstyle=o}
\psdots(5.4,9.1)(5.7,9.1)(6.0,9.4)(6.3,9.4)

%GRAPHIC 9a3
\psline{-}(7.5,8.8)(7.8,9.1) \psline{-}(7.8,9.1)(8.1,9.4)
\psline{-}(8.1,9.4)(8.4,9.7) \psline{-}(8.4,9.7)(8.7,9.4)
\psline{-}(8.7,9.4)(9.0,9.1) \psline{-}(9.0,9.1)(9.3,9.4)
\psline{-}(9.3,9.4)(9.6,9.7) \psline{-}(9.6,9.7)(9.9,9.4)
\psline{-}(9.9,9.4)(10.2,9.1) \psline{-}(10.2,9.1)(10.5,8.8)

%dots
\psset{dotsize=3pt}\psset{dotstyle=*}
\psdots(7.8,9.1)(8.1,9.4)(8.7,9.4)(9.3,9.4)(9.9,9.4)(10.2,9.1)

%GRAPHIC 9a4
\psline{-}(11.1,8.8)(11.4,9.1) \psline{-}(11.4,9.1)(11.7,9.4)
\psline{-}(11.7,9.4)(12.0,9.4) \psline{-}(12.0,9.4)(12.3,9.1)
\psline{-}(12.3,9.1)(12.6,9.1) \psline{-}(12.6,9.1)(12.9,9.4)
\psline{-}(12.9,9.4)(13.2,9.7) \psline{-}(13.2,9.7)(13.5,9.4)
\psline{-}(13.5,9.4)(13.8,9.1) \psline{-}(13.8,9.1)(14.1,8.8)

%dots
\psset{dotsize=3pt}\psset{dotstyle=*}
\psdots(11.4,9.1)(12.9,9.4)(13.5,9.4)(13.8,9.1)

\psset{dotsize=3pt}\psset{dotstyle=o}
\psdots(11.7,9.4)(12.0,9.4)(12.3,9.1)(12.6,9.1)

%GRAPHIC 9a5
\psline{-}(0.3,7.4)(0.6,7.7) \psline{-}(0.6,7.7)(0.9,8.0)
\psline{-}(0.9,8.0)(1.2,7.7) \psline{-}(1.2,7.7)(1.5,7.4)
\psline{-}(1.5,7.4)(1.8,7.7) \psline{-}(1.8,7.7)(2.1,8.0)
\psline{-}(2.1,8.0)(2.4,8.3) \psline{-}(2.4,8.3)(2.7,8.0)
\psline{-}(2.7,8.0)(3.0,7.7) \psline{-}(3.0,7.7)(3.3,7.4)

%dots
\psset{dotsize=3pt}\psset{dotstyle=*}
\psdots(0.6,7.7)(1.2,7.7)(1.8,7.7)(2.1,8.0)(2.7,8.0)(3.0,7.7)

%identification
\rput(-0.4,10.0){{\small $a)$}}
%\rput(2.8,10.0){{\small $(i,j)= (2,3)$}}

%GRAPHIC 9b1
\psline{-}(0.3,5.8)(0.6,6.1) \psline{-}(0.6,6.1)(0.9,6.4)
\psline{-}(0.9,6.4)(1.2,6.4) \psline{-}(1.2,6.4)(1.5,6.1)
\psline{-}(1.5,6.1)(1.8,5.8) \psline{-}(1.8,5.8)(2.1,6.1)
\psline{-}(2.1,6.1)(2.4,6.4) \psline{-}(2.4,6.4)(2.7,6.1)
\psline{-}(2.7,6.1)(3.0,5.8)

%dots
\psset{dotsize=3pt}\psset{dotstyle=*}
\psdots(0.6,6.1)(1.5,6.1)(2.1,6.1)(2.7,6.1)
\psset{dotsize=3pt}\psset{dotstyle=o}
\psdots(0.9,6.4)(1.2,6.4)

%GRAPHIC 9b2
\psline{-}(3.6,5.8)(3.9,6.1) \psline{-}(3.9,6.1)(4.2,6.4)
\psline{-}(4.2,6.4)(4.5,6.4) \psline{-}(4.5,6.4)(4.8,6.1)
\psline{-}(4.8,6.1)(5.1,6.1) \psline{-}(5.1,6.1)(5.4,6.4)
\psline{-}(5.4,6.4)(5.7,6.4) \psline{-}(5.7,6.4)(6.0,6.1)
\psline{-}(6.0,6.1)(6.3,5.8)

%dots
\psset{dotsize=3pt}\psset{dotstyle=*}
\psdots(3.9,6.1)(6.0,6.1)
\psset{dotsize=3pt}\psset{dotstyle=o}
\psdots(4.2,6.4)(4.5,6.4)(4.8,6.1)(5.1,6.1)(5.4,6.4)(5.7,6.4)

%GRAPHIC 9b3
\psline{-}(6.9,5.8)(7.2,6.1) \psline{-}(7.2,6.1)(7.5,6.4)
\psline{-}(7.5,6.4)(7.8,6.1) \psline{-}(7.8,6.1)(8.1,5.8)
\psline{-}(8.1,5.8)(8.4,6.1) \psline{-}(8.4,6.1)(8.7,6.4)
\psline{-}(8.7,6.4)(9.0,6.4) \psline{-}(9.0,6.4)(9.3,6.1)
\psline{-}(9.3,6.1)(9.6,5.8)

%dots
\psset{dotsize=3pt}\psset{dotstyle=*}
\psdots(7.2,6.1)(7.8,6.1)(8.4,6.1)(9.3,6.1)
\psset{dotsize=3pt}\psset{dotstyle=o}
\psdots(8.7,6.4)(9.0,6.4)

%identification
\rput(-0.4,6.7){{\small $b)$}}
%\rput(2.8,6.7){{\small $(i,j)= (2,5/2)$}}

%GRAPHIC 9c1
\psline{-}(0.3,3.9)(0.6,4.2) \psline{-}(0.6,4.2)(0.9,4.5)
\psline{-}(0.9,4.5)(1.2,4.8) \psline{-}(1.2,4.8)(1.5,4.5)
\psline{-}(1.5,4.5)(1.8,4.2) \psline{-}(1.8,4.2)(2.1,3.9)
\psline{-}(2.1,3.9)(2.4,4.2) \psline{-}(2.4,4.2)(2.7,4.2)
\psline{-}(2.7,4.2)(3.0,3.9)

%dots
\psset{dotsize=3pt}\psset{dotstyle=*}
\psdots(0.6,4.2)(0.9,4.5)(1.5,4.5)(1.8,4.2)
\psset{dotsize=3pt}\psset{dotstyle=o}
\psdots(2.4,4.2)(2.7,4.2)

%GRAPHIC 9c2
\psline{-}(3.6,3.9)(3.9,4.2) \psline{-}(3.9,4.2)(4.2,4.5)
\psline{-}(4.2,4.5)(4.5,4.8) \psline{-}(4.5,4.8)(4.8,4.5)
\psline{-}(4.8,4.5)(5.1,4.2) \psline{-}(5.1,4.2)(5.4,4.2)
\psline{-}(5.4,4.2)(5.7,4.5) \psline{-}(5.7,4.5)(6.0,4.2)
\psline{-}(6.0,4.2)(6.3,3.9)

%dots
\psset{dotsize=3pt}\psset{dotstyle=*}
\psdots(3.9,4.2)(4.2,4.5)(4.8,4.5)(6.0,4.2)
\psset{dotsize=3pt}\psset{dotstyle=o}
\psdots(5.1,4.2)(5.4,4.2)

%GRAPHIC 9c3
\psline{-}(7.2,3.9)(7.5,4.2) \psline{-}(7.5,4.2)(7.8,4.5)
\psline{-}(7.8,4.5)(8.1,4.8) \psline{-}(8.1,4.8)(8.4,4.5)
\psline{-}(8.4,4.5)(8.7,4.2) \psline{-}(8.7,4.2)(9.0,4.5)
\psline{-}(9.0,4.5)(9.3,4.5) \psline{-}(9.3,4.5)(9.6,4.2)
\psline{-}(9.6,4.2)(9.9,3.9)

%dots
\psset{dotsize=3pt}\psset{dotstyle=*}
\psdots(7.5,4.2)(7.8,4.5)(8.4,4.5)(9.6,4.2)
\psset{dotsize=3pt}\psset{dotstyle=o}
\psdots(9.0,4.5)(9.3,4.5)

%GRAPHIC 9c4
\psline{-}(10.5,3.9)(10.8,4.2) \psline{-}(10.8,4.2)(11.1,4.5)
\psline{-}(11.1,4.5)(11.4,4.8) \psline{-}(11.4,4.8)(11.7,4.5)
\psline{-}(11.7,4.5)(12.0,4.5) \psline{-}(12.0,4.5)(12.3,4.8)
\psline{-}(12.3,4.8)(12.6,4.5) \psline{-}(12.6,4.5)(12.9,4.2)
\psline{-}(12.9,4.2)(13.2,3.9)

%dots
\psset{dotsize=3pt}\psset{dotstyle=*}
\psdots(10.8,4.2)(11.1,4.5)(12.6,4.5)(12.9,4.2)
\psset{dotsize=3pt}\psset{dotstyle=o}
\psdots(11.7,4.5)(12.0,4.5)

%GRAPHIC 9c5
\psline{-}(0.3,2.7)(0.6,3.0) \psline{-}(0.6,3.0)(0.9,3.3)
\psline{-}(0.9,3.3)(1.2,3.3) \psline{-}(1.2,3.3)(1.5,3.0)
\psline{-}(1.5,3.0)(1.8,3.3) \psline{-}(1.8,3.3)(2.1,3.6)
\psline{-}(2.1,3.6)(2.4,3.3) \psline{-}(2.4,3.3)(2.7,3.0)
\psline{-}(2.7,3.0)(3.0,2.7)

%dots
\psset{dotsize=3pt}\psset{dotstyle=*}
\psdots(0.6,3.0)(1.8,3.3)(2.4,3.3)(2.7,3.0)
\psset{dotsize=3pt}\psset{dotstyle=o}
\psdots(0.9,3.3)(1.2,3.3)

%GRAPHIC 9c6
\psline{-}(3.6,2.7)(3.9,3.0) \psline{-}(3.9,3.0)(4.2,3.3)
\psline{-}(4.2,3.3)(4.5,3.0) \psline{-}(4.5,3.0)(4.8,3.0)
\psline{-}(4.8,3.0)(5.1,3.3) \psline{-}(5.1,3.3)(5.4,3.6)
\psline{-}(5.4,3.6)(5.7,3.3) \psline{-}(5.7,3.3)(6.0,3.0)
\psline{-}(6.0,3.0)(6.3,2.7)

%dots
\psset{dotsize=3pt}\psset{dotstyle=*}
\psdots(3.9,3.0)(5.1,3.3)(5.7,3.3)(6.0,3.0)
\psset{dotsize=3pt}\psset{dotstyle=o}
\psdots(4.5,3.0)(4.8,3.0)
%identification

%GRAPHIC 9c7
\psline{-}(7.2,2.7)(7.5,3.0) \psline{-}(7.5,3.0)(7.8,3.0)
\psline{-}(7.8,3.0)(8.1,2.7) \psline{-}(8.1,2.7)(8.4,3.0)
\psline{-}(8.4,3.0)(8.7,3.3) \psline{-}(8.7,3.3)(9.0,3.6)
\psline{-}(9.0,3.6)(9.3,3.3) \psline{-}(9.3,3.3)(9.6,3.0)
\psline{-}(9.6,3.0)(9.9,2.7)

%dots
\psset{dotsize=3pt}\psset{dotstyle=*}
\psdots(8.4,3.0)(8.7,3.3)(9.3,3.3)(9.6,3.0)
\psset{dotsize=3pt}\psset{dotstyle=o}
\psdots(7.5,3.0)(7.8,3.0)

\rput(-0.4,5.1){{\small $c)$}}
%\rput(2.8,5.1){{\small $(i,j)= (3/2,3)$}}

%GRAPHIC 9d1
\psline{-}(0.3,1.1)(0.6,1.4) \psline{-}(0.6,1.4)(0.9,1.7)
\psline{-}(0.9,1.7)(1.2,1.7) \psline{-}(1.2,1.7)(1.5,1.4)
\psline{-}(1.5,1.4)(1.8,1.1) \psline{-}(1.8,1.1)(2.1,1.4)
\psline{-}(2.1,1.4)(2.4,1.4) \psline{-}(2.4,1.4)(2.7,1.1)

%dots
\psset{dotsize=3pt}\psset{dotstyle=*}
\psdots(0.6,1.4)(1.5,1.4)
\psset{dotsize=3pt}\psset{dotstyle=o}
\psdots(0.9,1.7)(1.2,1.7)(2.1,1.4)(2.4,1.4)

%GRAPHIC 9d2
\psline{-}(3.3,1.1)(3.6,1.4) \psline{-}(3.6,1.4)(3.9,1.7)
\psline{-}(3.9,1.7)(4.2,1.7) \psline{-}(4.2,1.7)(4.5,1.4)
\psline{-}(4.5,1.4)(4.8,1.4) \psline{-}(4.8,1.4)(5.1,1.7)
\psline{-}(5.1,1.7)(5.4,1.4) \psline{-}(5.4,1.4)(5.7,1.1)

%dots
\psset{dotsize=3pt}\psset{dotstyle=*}
\psdots(3.6,1.4)(5.4,1.4)
\psset{dotsize=3pt}\psset{dotstyle=o}
\psdots(3.9,1.7)(4.2,1.7)(4.5,1.4)(4.8,1.4)

%GRAPHIC 9d3
\psline{-}(6.3,1.1)(6.6,1.4) \psline{-}(6.6,1.4)(6.9,1.7)
\psline{-}(6.9,1.7)(7.2,1.7) \psline{-}(7.2,1.7)(7.5,1.4)
\psline{-}(7.5,1.4)(7.8,1.7) \psline{-}(7.8,1.7)(8.1,1.7)
\psline{-}(8.1,1.7)(8.4,1.4) \psline{-}(8.4,1.4)(8.7,1.1)

%dots
\psset{dotsize=3pt}\psset{dotstyle=*}
\psdots(6.6,1.4)(8.4,1.4)
\psset{dotsize=3pt}\psset{dotstyle=o}
\psdots(6.9,1.7)(7.2,1.7)(7.8,1.7)(8.1,1.7)

%GRAPHIC 9d4
\psline{-}(9.3,1.1)(9.6,1.4) \psline{-}(9.6,1.4)(9.9,1.7)
\psline{-}(9.9,1.7)(10.2,1.4) \psline{-}(10.2,1.4)(10.5,1.4)
\psline{-}(10.5,1.4)(10.8,1.7) \psline{-}(10.8,1.7)(11.1,1.7)
\psline{-}(11.1,1.7)(11.4,1.4) \psline{-}(11.4,1.4)(11.7,1.1)

%dots
\psset{dotsize=3pt}\psset{dotstyle=*}
\psdots(9.6,1.4)(11.4,1.4)
\psset{dotsize=3pt}\psset{dotstyle=o}
\psdots(10.2,1.4)(10.5,1.4)(10.8,1.7)(11.1,1.7)

%GRAPHIC 9d5
\psline{-}(0.3,0.0)(0.6,0.3) \psline{-}(0.6,0.3)(0.9,0.3)
\psline{-}(0.9,0.3)(1.2,0.0) \psline{-}(1.2,0.0)(1.5,0.3)
\psline{-}(1.5,0.3)(1.8,0.6) \psline{-}(1.8,0.6)(2.1,0.6)
\psline{-}(2.1,0.6)(2.4,0.3) \psline{-}(2.4,0.3)(2.7,0.0)

%dots
\psset{dotsize=3pt}\psset{dotstyle=*}
\psdots(1.5,0.3)(2.4,0.3)
\psset{dotsize=3pt}\psset{dotstyle=o}
\psdots(0.6,0.3)(0.9,0.3)(1.8,0.6)(2.1,0.6)

%identification
\rput(-0.4,2.0){{\small $d)$}}
%\rput(2.8,2.0){{\small $(i,j)= (3/2,5/2)$}}

\end{pspicture}
\end{center}
\end{figure}
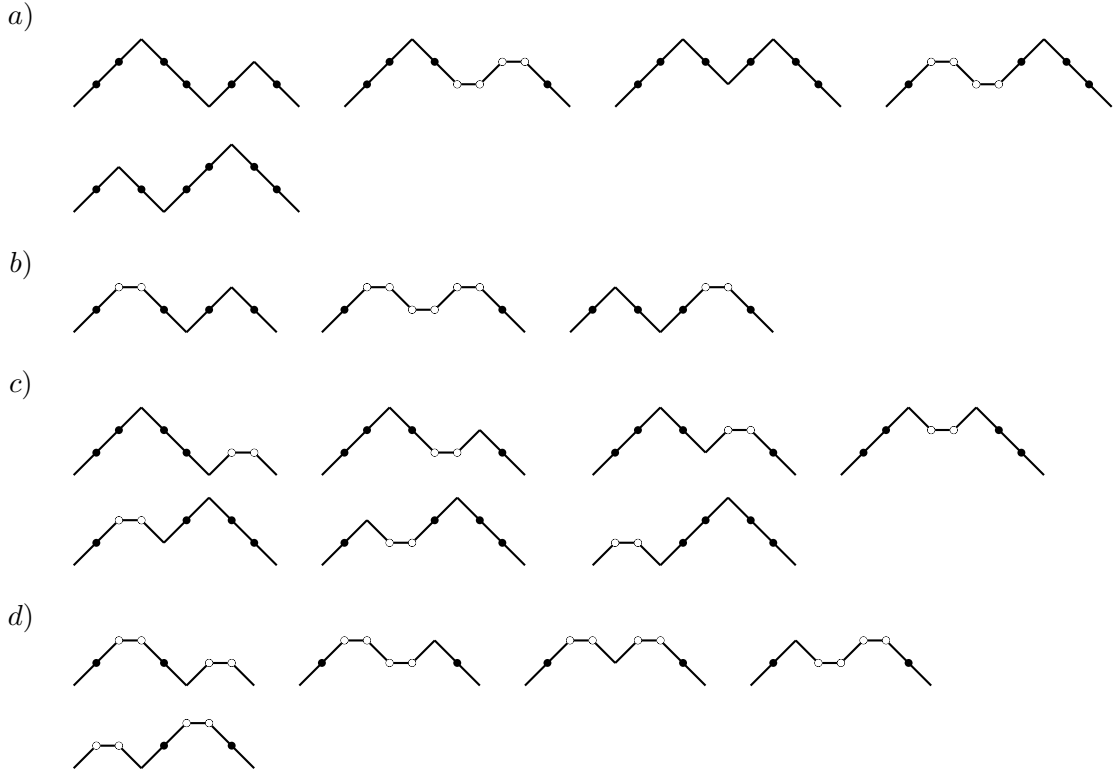 
%%%%%%%%%%%%%%%%%%
%Note that given a symmetrical configuration, it is the leftmost peak which is ascribed the largest charge. 

 % 
 
% \n [Expliquer comment lire le contenu en charge d'un complexe a partir de ces regles: testons ceci:]
 
 A complex refers to a part of the  path that is delimitated by two points on the horizontal axis (that is, with height 1).
 The charge of a peak at position $i$ and height $h$  is (up to possible contributions of one or two H edges in a way to be explained below) the largest number $c\geq 1$ such that we can find two vertices $(i',h-c)$ and $(i'',h-c)$ on the  path  with $i'<i<i''$ and such that between these two vertices there are no peak of height larger than $h$ and every peak of height equal to $h$ is located at its right \cite{BP}. Draw a baseline at vertical distance $c$ from the peak.  If  there are H edges on the baseline which are preceded or followed by a  peak with height  larger than the one under study, add 
an extra 1/2 to the height. If there are H edges on the top of the peak, add 1/2 to the charge. Delimitate the particles using their charge determination, that is, identify those H edges that are part of the particles of charge $\geq 3/2$. Take the rule that there can be no free (i.e, which is not part of a particle of charge $\geq 3/2$) H edge in-between particles. The remaining H edges are interpreted as particles of charge 1/2. For instance, the charge content of the path in Fig. \ref{fig3} is detailed in the figure caption.

 \subsection{Strategy for constructing the generating function}

The construction of the generating function proceeds  in various steps \cite{OleJS} which are detailed in the following subsections:

\begin{enumerate}

\item For a fixed particle content, identify the ordering of the peaks that minimizes the weight %configuration with lowest weight 
and evaluate the  weight of this minimal-weight configuration. With $n_j$ denoting the number of particles of charge $j$, a fixed particle content means a fixed set of values $\{n_j\}$.

\item  Identify all possible ways  of modifying  the ordering of the peaks of charge $\geq 3/2$ from their position in the  minimal-weight configuration and determine their weight relative to that  of  the minimal-weight configuration.

\item Identify all possible displacements of the particles of charges 1/2 and 1 within the sequence of particles of  charge $\geq 3/2$ (whose ordering is kept fixed)  and determine the corresponding  weight.

%\item In each configuration just identified, determine all possible displacements of the peaks of charge 1/2 and 1 within the sequence of particles of  charge $\geq 3/2$ (whose ordering is kept fixed)  and determine the corresponding  weight.

\item The first three items yield the full data necessary to construct the generating function for all paths with a fixed particle content. In order to get the full generating function, it remains to sum over all possible charges $n_j$ compatible with the fixed length
 $L$, with 
 \begin{equation}
 L= 2{\sum_{j=1/2}^{k/2+1}} \,j\,  n_{j} ,
 \end{equation}
where  the summation is incremented by steps of 1/2 (and it will always be clear from the context whenever this is the case, in sums or products).
 
\end{enumerate}

% But before we consider these points, it is appropriate to analyze briefly the gross features of the particle dynamics.

\subsection{The minimal-weight configuration}

The configuration of minimal weight with specified and fixed value of the $n_j$ is the following: all the particles of charge $\geq 3/2$ are ordered in decreasing values of the charge, from right to left;  all the H links describing the various particles of charge 1/2 are inserted in rightmost position  within the last particle of lowest charge, which is generically a particle of charge $3/2$. Then the path is terminated by the sequence of particles of charge 1. An example is displayed in Fig. \ref{fig6}.

%%%%%%%%%%%%%%%%%%%%%%
%%%% Figure 6 Fermionic Gaz Susy                          %%%%
%%%%%%%%%%%%%%%%%%%%%%%%%%%%%%%%%%%%%%%%%%%%%%%%%%%%%%%%%%%%%%

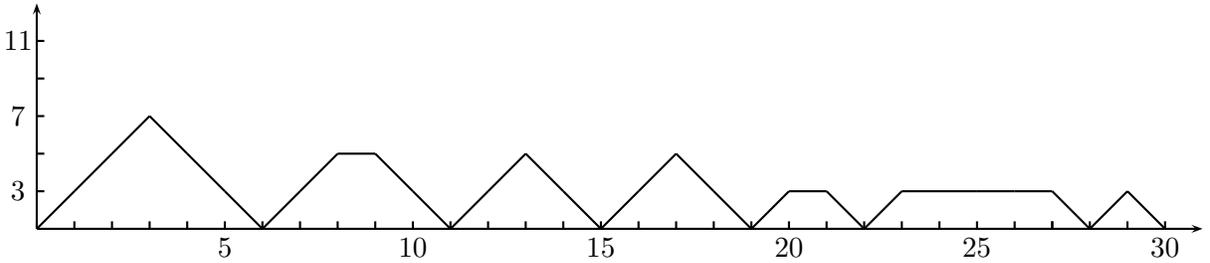
\begin{figure}[ht]
\caption{{\footnotesize The minimal-weight configuration with particle content: $n_3 =n_{\frac52}=1$, $n_2 = n_{\frac32}=2$, $n_1=1$ and $n_{\frac12}=3$. Note that among the four H edges  between 24 and 27, the first one is a constituent edge of the charge 3/2 particle.}}
\label{fig6}
\vskip0,4cm
\begin{center}
\begin{pspicture}(0,0)(15.5,3.6)
%axis
\psline{->}(0.5,0.5)(0.5,3.5) \psline{->}(0.5,0.5)(16.0,0.5)
%\psset{linestyle=dashed,dashadjust=false} \psline(0.5,2.0)(15.5,2.0)
%\psset{linestyle=dotted}
%\psset{linestyle=solid}
%\psline{<->}(4.5,1.0)(6.5,1.0)
\psset{linestyle=solid}
%units
\psline{-}(0.5,0.5)(0.5,0.6) \psline{-}(1.0,0.5)(1.0,0.6)
\psline{-}(1.5,0.5)(1.5,0.6) \psline{-}(2.0,0.5)(2.0,0.6)
\psline{-}(2.5,0.5)(2.5,0.6) \psline{-}(3.0,0.5)(3.0,0.6)
\psline{-}(3.5,0.5)(3.5,0.6) \psline{-}(4.0,0.5)(4.0,0.6)
\psline{-}(4.5,0.5)(4.5,0.6) \psline{-}(5.0,0.5)(5.0,0.6)
\psline{-}(5.5,0.5)(5.5,0.6) \psline{-}(6.0,0.5)(6.0,0.6)
\psline{-}(6.5,0.5)(6.5,0.6) \psline{-}(7.0,0.5)(7.0,0.6)
\psline{-}(7.5,0.5)(7.5,0.6) \psline{-}(8.0,0.5)(8.0,0.6)
\psline{-}(8.5,0.5)(8.5,0.6) \psline{-}(9.0,0.5)(9.0,0.6)
\psline{-}(9.5,0.5)(9.5,0.6) \psline{-}(10.0,0.5)(10.0,0.6)
\psline{-}(10.5,0.5)(10.5,0.6) \psline{-}(11.0,0.5)(11.0,0.6)
\psline{-}(11.5,0.5)(11.5,0.6) \psline{-}(12.0,0.5)(12.0,0.6)
\psline{-}(12.5,0.5)(12.5,0.6) \psline{-}(13.0,0.5)(13.0,0.6)
\psline{-}(13.5,0.5)(13.5,0.6) \psline{-}(14.0,0.5)(14.0,0.6)
\psline{-}(14.0,0.5)(14.0,0.6) \psline{-}(14.5,0.5)(14.5,0.6)
\psline{-}(15.0,0.5)(15.0,0.6) \psline{-}(15.5,0.5)(15.5,0.6)
%\psline{-}(15.0,0.5)(15.0,0.6) \psline{-}(15.5,0.5)(15.5,0.6)
\rput(3.0,0.25){{\small $5$}} 
\rput(5.5,0.25){{\small $10$}} \rput(8.0,0.25){{\small $15$}}
\rput(10.5,0.25){{\small $20$}} \rput(13.0,0.25){{\small $25$}}
\rput(15.5,0.25){{\small $30$}}
 \psline{-}(0.5,1.0)(0.6,1.0)
\psline{-}(0.5,1.5)(0.6,1.5) \psline{-}(0.5,2.0)(0.6,2.0)
\psline{-}(0.5,2.5)(0.6,2.5) \psline{-}(0.5,3.0)(0.6,3.0)
\rput(0.25,1.0){{\small $3$}} \rput(0.25,2.0){{\small $7$}}
 \rput(0.25,3.0){{\small $11$}}
%graphic
\psline{-}(0.5,0.5)(1.0,1.0) \psline{-}(1.0,1.0)(1.5,1.5)
\psline{-}(1.5,1.5)(2.0,2.0) \psline{-}(2.0,2.0)(2.5,1.5)
\psline{-}(2.5,1.5)(3.0,1.0) \psline{-}(3.0,1.0)(3.5,0.5)
\psline{-}(3.5,0.5)(4.0,1.0) \psline{-}(4.0,1.0)(4.5,1.5)
\psline{-}(4.5,1.5)(5.0,1.5) \psline{-}(5.0,1.5)(5.5,1.0)
\psline{-}(5.5,1.0)(6.0,0.5) \psline{-}(6.0,0.5)(6.5,1.0)
\psline{-}(6.5,1.0)(7.0,1.5) \psline{-}(7.0,1.5)(7.5,1.0)
\psline{-}(7.5,1.0)(8.0,0.5) \psline{-}(8.0,0.5)(8.5,1.0)
\psline{-}(8.5,1.0)(9.0,1.5) \psline{-}(9.0,1.5)(9.5,1.0)
\psline{-}(9.5,1.0)(10.0,0.5)
 \psline{-}(10.0,0.5)(10.5,1.0)\psline{-}(10.5,1.0)(11.0,1.0)
\psline{-}(11.0,1.0)(11.5,0.5) \psline{-}(11.5,0.5)(12.0,1.0)
\psline{-}(12.0,1.0)(12.5,1.0) \psline{-}(12.5,1.0)(13.0,1.0)
\psline{-}(13.0,1.0)(13.5,1.0) \psline{-}(13.5,1.0)(14.0,1.0)
\psline{-}(14.0,1.0)(14.5,0.5) \psline{-}(14.5,0.5)(15.0,1.0)
\psline{-}(15.0,1.0)(15.5,0.5) 
%\psline{-}(15.5,0.5)(16.0,1.0)\psline{-}(16.0,1.0)(16.5,0.5) 
%dots
%\psset{dotsize=3pt}
%\psdots(2.5,1.0)(3.0,1.5)(4.0,1.5)(5.0,1.5)(6.0,1.5)(6.5,1.0)(10.0,1.0)(11.0,1.0)

\end{pspicture}
\end{center}
\end{figure}

Our first objective is to evaluate the weight of this configuration. We first compute the weight of an isolated particle of integer charge $r$ and then a sequence of these.
For a peak centered on the point $r+x_0$, the weight is 
 \begin{equation}\sum_{j=1}^{2r-1} (j+x_0)-(r+x_0)= 2(r-1)(r+x_0).
 \end{equation}
 In this expression, we have summed the contribution of all vertices in-between $x_0$ and $2r+x_0$ (recall that $2r$ is the diameter of the particle) and subtract the non-contributing middle point, the position of the maximum. 
 Consider next the contribution of $n_r$ adjacent  particles of charge $r$:
  \begin{equation}\label{tot}
  \sum_{j=1}^{2rn_r} (j+x_0)-\sum_{j=1}^{2n_r}(jr+x_0)= 2(r-1)n_r(rn_r+x_0).
 \end{equation}
 Here we add all integer points $x$ such that $x_0<x\leq x_0+2rn_r$ and remove those points that correspond to extrema, at positions $jr+x_0$.
 In the minimal-weight configuration, these peaks of charge $r$ are preceded by the sequence of all higher charge peaks. This fixes the value of $x_0$ to 
   \begin{equation} x_0 =\sum_{j=r+1/2}^{k/2+1} 2j n_j.
 \end{equation}

 Consider next the contribution of a particle of half-integer charge $r$ centered at the position $\lfloor r\rfloor+x_0+1/2$.  
It is described by  a flatten triangle of height $ \lfloor r\rfloor$. The weight is 
 \begin{equation}\sum_{j=1}^{2r-1} (j+x_0)-(\lfloor r\rfloor+x_0)- (\lfloor r\rfloor+1+x_0) + \frac12(\lfloor r\rfloor+x_0) +\frac12(\lfloor r\rfloor+1+x_0)  = 2(r-1)(r+x_0).
 \end{equation}
 Here we have subtracted the contributions of the two top corners of the flatten triangle from the sum and then added their contribution to the weight (which is half that of the other points) separately. Since 
$ 2\lfloor r\rfloor+1=2r$, we end up with the same expression as in the integer charge case. The weight of $n_r$ peaks is also given by (\ref{tot}).

 Putting these results together, the energy of the ordered sequence of particles of charge $\geq 3/2$ is found to be 
 \begin{equation}
 \sum_{i,j=3/2}^{k/2+1} n_i B_{ij} n_j ,
  \end{equation}
  where $B$ is the  matrix whose  entries are given by
 \begin{equation}\label{Bij}
B_{ij}=B_{ji} \qquad \text{and}\qquad B_{ij} = 2(i-1)j \qquad \text{for}\qquad  i \leq j.
\end{equation}
The insertion of the H edges describing the $n_{1/2}$ particles of charge 1/2 displaces the SE edge of the last particle (whose charge is the lowest charge $\geq 3/2$  in the configuration) by $n_{1/2} $ units. This increases the weight by $(n_{1/2})/2$. Finally, the particles of charge 1 appended to the end of the path do not contribute to the weight.

The weight of the minimal-weight configuration $w_{\rm mwc}$ with fixed particle content, that is, fixed values of all $n_j$, is thus
 \begin{equation}
w_{\rm mwc}=  \sum_{i,j=3/2}^{k/2+1} n_i B_{ij} n_j +\frac12 n_{1/2}.
\end{equation}

\subsection{Mixing the higher charge particles among themselves}

%Up to this point, we have only considered the insertion of particle of charges 1/2 and 1 within the larger ones. This is equivalent to displacing the larger particles (of charges $\geq 3/2$) in all possible ways toward the left but still  maintaining their respective ordering in non-increasing values of the charge (from left to right). In that way, a huge number of allowed configurations have still not been considered, namely those for which these larger particles are mixed among themselves.
% is thus not affected. 

%The combinatorial description of these cases is rather similar to the situations already treated. 

Starting from the minimal-weight configuration, we now consider all possible successive displacements of the particles of charge $k/2, \cdots, 2, 3/2$ within the larger ones. 
In this process it is understood that the length filling  particles of charge 1 and 1/2 remain fixed. In particular, this means that the charge 1/2 particles remain at the same horizontal and vertical positions: they are kept attached to the rightmost particle of charge $\geq 3/2$ (irrespective of its charge once the mixing is completed).
% and at the precise height they had in the minimal weight configuration.

The first step is to determine the number of ways a particle of charge $i$ can be inserted in a particle of charge $j>i$. 
The  four possible parities of $2i$ and $2j$ need to be  treated separately but in all cases we find that the number of configurations is $4(j-i)+1$. 

Consider first the case where $i$ and $j$ are both integers. In the light of  the deformation process described previously, there are generically two allowed configurations at each insertion point.
There are $(j-i)$ such points on the straight-down side of the particle of charge $j$ and $(j-i)$ in its straight-up side. However, in the latter case, at the  insertion point closest to the top of the large peak, only one configuration is distinct. This gives a total of $4(j-i)-1$. To this number, we add the configuration where the particle of charge $i$ is either before or after the larger one, for a total of $4(j-i)+1$. The same counting holds when $i$ and $j$ are both half-integer.
For $i$ integer and $j$ half-integer, we have a total of $2(\lf j\rf-i)+1+1$ configurations where the initial vertex of the particle $i$ is inserted in the straight-down part of the larger particle and $2(\lf j\rf-i-1)+1$  configurations where the last vertex of the particle $i$ is inserted in its straight-up portion. For the reversed parities, the numbers are respectively $2(j-\lc i\rc)+1+1$ and  $2(j-\lc i\rc)+1$. (The cases where the particles are actually separated are taken into account with the final  $+1$ in each expressions). In both cases, the total is $4(j-i)+1$. All cases are illustrated in Fig. \ref{fig5}.
For $n_i$ and $n_j$ particles of each type, this is generalized to the combinatorial factor
\begin{equation}\label{bino}
\begin{pmatrix} n_{i}+4(j-i)n_j\\ n_i\end{pmatrix}.
\end{equation} 
%$n_i+4(j-i)n_j$.

It remains to weight the different steps.  We start from the initial configuration where the two particles are separated, with $j$  at  the left of $i$ 
 (which is their relative position in the minimal-weight configuration). The starting distance between the particles is thus  $j+i$. The successive configurations describe the interpenetration of $i$ within $j$ such that, at each step, the separation distance decreases by $1/2$. This is pursued until their minimal separation is reached. From there on, the particle identities are interchanged and the subsequent displacements are performed from the leftmost peak, so that the distance increases by 1/2 at each step.  Each such displacement induces a weight difference of 1. This is illustrated in Fig. \ref{fig5} for number of special cases. 
 
 In order to demonstrate this in general, one compares two successive configurations (with particle separation differing by 1/2) and observes that, for both parities of $2i$, only four vertices are modified in their weight contribution. For a penetration in the straight-down part of the larger particle, the different possibilities are pictured in Fig. \ref{fig10} for the case of $i$ integer. The four vertices whose weight are affected are diamond-shaped. By comparing the total weight of these four vertices in the two successive configurations, one readily  verifies that the difference is 1 in each case (cf. the sample calculation in the figure caption). The net effect of this weight change is that the binomial factor (\ref{bino}) is simply $q$-deformed:
 \begin{equation}\label{binoi}
\begin{bmatrix} n_{i}+4(j-i)n_j\\ n_i\end{bmatrix},
\end{equation} 
where
 \begin{equation}
\begin{bmatrix}
a\\ b\end{bmatrix}_q =  \begin{cases} & \frac{(q)_a}{(q)_{a-b}(q)_b}  \quad \text{if}\quad 0\leq b\leq a,\\ &\qquad 0\qquad\qquad\text{otherwise},
\end{cases}
\end{equation} 
with
\begin{equation}\label{zqde}
\qquad (z)_a\equiv (z;q)_a= (1-z)\cdots (1-zq^{a-1}).
\end{equation}

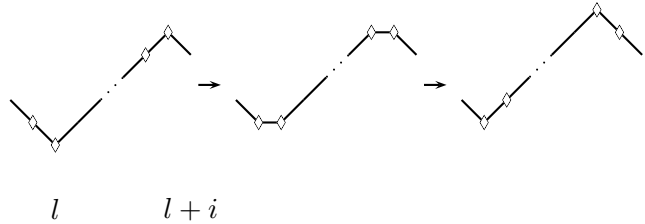
\begin{figure}[ht]
\caption{{\footnotesize Three successive configurations describing the interpenetration of a particle of integer charge $i$ inside a larger particle. Here only four vertices are displayed and their horizontal positions, $l-1, \, l,\, l+i-1,\,l+i$, are unchanged from one configuration to the other. Only these  four indicated vertices  have their weight modified in these processes. The weight of the four vertices in these three configurations is respectively $(l-1,0,l+i-1,0)$, $\frac12(l-1,l,l+i-1,l+i)$ and $(0,l,0,l+i)$. The sums of the corresponding four weights are  $2l+i-2,\ 2l+i-1$ and $2l+i$. The difference is thus 1 at each step. The figure for the case where $i$ is half-integer is similar but the four vertices whose weight is changed
are not the same in the two steps.
% in the first step are not the same as those in the second step.
}} \label{fig10}
\begin{center}
\begin{pspicture}(0,0)(9.0,3.0)

%GRAPHIC 11a
\psline{-}(0.3,1.7)(0.6,1.4) \psline{-}(0.6,1.4)(0.9,1.1)
\psline{-}(0.9,1.1)(1.2,1.4) \psline{-}(1.2,1.4)(1.5,1.7)
\psset{linestyle=dotted}
\psline{-}(1.5,1.7)(1.8,2.0)
\psset{linestyle=solid}
 \psline{-}(1.8,2.0)(2.1,2.3)
\psline{-}(2.1,2.3)(2.4,2.6) \psline{-}(2.4,2.6)(2.7,2.3)

%dots
\psset{dotsize=4pt}\psset{dotstyle=diamond}
\psdots(0.6,1.4)(0.9,1.1)(2.1,2.3)(2.4,2.6)

 \psline{->}(2.8,1.9)(3.1,1.9)

%GRAPHIC 11b
\psline{-}(3.3,1.7)(3.6,1.4) \psline{-}(3.6,1.4)(3.9,1.4)
\psline{-}(3.9,1.4)(4.2,1.7) \psline{-}(4.2,1.7)(4.5,2.0)
\psset{linestyle=dotted}
\psline{-}(4.5,2.0)(4.8,2.3)
\psset{linestyle=solid}
 \psline{-}(4.8,2.3)(5.1,2.6)
\psline{-}(5.1,2.6)(5.4,2.6) \psline{-}(5.4,2.6)(5.7,2.3)

%dots
\psset{dotsize=4pt}\psset{dotstyle=diamond}
\psdots(3.6,1.4)(3.9,1.4)(5.1,2.6)(5.4,2.6)

 \psline{->}(5.8,1.9)(6.1,1.9)

%GRAPHIC 11c
\psline{-}(6.3,1.7)(6.6,1.4) \psline{-}(6.6,1.4)(6.9,1.7)
\psline{-}(6.9,1.7)(7.2,2.0) 
\psset{linestyle=dotted}
\psline{-}(7.2,2.0)(7.5,2.3)
\psset{linestyle=solid}
\psline{-}(7.5,2.3)(7.8,2.6) \psline{-}(7.8,2.6)(8.1,2.9)
\psline{-}(8.1,2.9)(8.4,2.6) \psline{-}(8.4,2.6)(8.7,2.3)

%dots
\psset{dotsize=4pt}\psset{dotstyle=diamond}
\psdots(6.6,1.4)(6.9,1.7)(8.1,2.9)(8.4,2.6)

%text
\rput(0.9,0.25){{\small $l$}} \rput(2.7,0.25){{\small $l+i$}}
%\rput(3.9,0.25){{\small $l$}} \rput(5.7,0.25){{\small $l+i$}}
%\rput(6.9,0.25){{\small $l-1$}} \rput(8.7,0.25){{\small $l+i-1$}}

\end{pspicture}
\end{center}
\end{figure}

 %the various steps of penetration iinvlove

The generalization to more complex configurations is immediate. The whole number  of $q$-weighted configurations is thus given by 
\begin{equation}
 \prod_{i=3/2}^{k/2+1}
\begin{bmatrix}
n_{i}+m_i  \\ n_{i}\end{bmatrix}_q
\end{equation}
where 
\begin{equation}\label{mi}
m_i=\sum_{j=i+1/2}^{k/2+1} 4(j-i)n_{j}\qquad (i>1).
\end{equation}

\subsection{Mixing the particles of charge  1 }

The $q$-weighted combinatorial factor accounting for the different possible insertions of the $n_{1}$ particles of charge 1 within those of higher charge, with the charged 1/2 particle still maintained fixed, is given by the $q^2$-binomial factor
\begin{equation}
\begin{bmatrix}
n_{1}+M  \\ n_{1}\end{bmatrix}_{q^2},
\end{equation}
%with $M$ defined in (\ref{m1}).
with
$M$ defined by
\begin{equation}\label{m1} M= \sum_{j=1}^{k} jn_{\frac{j}2+1}. \end{equation}
The analysis is similar to that of the previous subsection. The main difference
 is that a charge 1 particle cannot be deformed: this affects both the combinatorics and the weight change. 
% and this means that  the separation distance between two penetration steps is 1. 

Consider first the number of insertions  of a particle of charge 1 within a particle of charge $j$. For $j$ integer, there are $2j-3$ insertion points strictly within the particle (eliminating one case that would lead to a  configuration already considered) plus 2 configurations corresponding to  the particle of charge 1 being outside the larger one, either behind or before it. The total is $2(j-1)+1$ possibilities. For $j$ half-integer, the counting is similar: there are $2\lf j\rf -2 $ insertion points within the particle plus the two outside, again for a total of $2(j-1)+1$ possibilities.
In both cases (both parities of $2j$), for $n_j$ particles of type $j$, there are  $2(j-1)n_j+1$ distinct configurations and when there are $n_1$ particles of charge 1, this becomes 
\begin{equation}\label{com1}
\begin{pmatrix} n_{1}+2(j-1)n_j\\ n_1\end{pmatrix}.
\end{equation}  
Again, for a more general charge content, the factor $2(j-1) n_j$ is changed into $M$.

It remains to check that every move toward the left of a particle of charge 1 within a sequence of larger particles induces a weight increase of 2. This operation amounts to displace by two units toward the right either a (NE,NE) or a (SE,SE) vertex, which indeed augments the weight by 2. This is illustrated in Fig. \ref{fig7}, for the insertion of a charge
1 into a charge 5/2 particle. The result also follows from the adaptation of  the analysis of the preceding subsection with $i=1$. However, because the charge 1 particle is not deformable, the intermediate configurations that previously accounted for a weight difference of 1 are missing, so that 
every allowed move changes the weight by 2. As a result, (\ref{com1}) is not $q$-deformed but rather $q^2$-deformed.
%The number of resulting configurations is roughly the same as in the previous case: a particle of charge 1 can be inserted at $2(j-1)$ positions within the particle of charge $>1$. This includes the configurations where the charge 1 particle is either in front of behind the larger particle. When compared with the number of configurations obtained by inserting the charge 1/2 particles, we see that these two additional configurations compensate for the two configurations excluded by that the fact that the minimal distance are now 2 or 3 depending of the relative positions of the particle 1.  Note also that  the  absence of the factor of $-1$: we start with a configuration where the charge 1 particles lie outside the larger particles. A more important  difference between the two cases is that the change of weight resulting from each displacement of a particle of charge 1 toward the left  nduces an increase in the weight equal 2. (faire un ex et donner l'argument general). This is the origin of the $q^2$-binomial factor.

\begin{figure}[ht]
\caption{{\footnotesize The interpenetration of a particle of charge $1$ through a particle of charge $5/2$. The weight of the successive configurations differs by 2. This is easily seen here since either a black dot is moved by two units toward the right (as in the first and third steps) or two circled dots are similarly moved by one units each (as in the second step. }} \label{fig7}
\begin{center}
\begin{pspicture}(0,0)(11.0,2.0)

%GRAPHIC 8a
\psline{-}(0.3,1.1)(0.6,1.4) \psline{-}(0.6,1.4)(0.9,1.7)
\psline{-}(0.9,1.7)(1.2,1.7) \psline{-}(1.2,1.7)(1.5,1.4)
\psline{-}(1.5,1.4)(1.8,1.1) \psline{-}(1.8,1.1)(2.1,1.4)
\psline{-}(2.1,1.4)(2.4,1.1)

%dots
\psset{dotsize=3pt}\psset{dotstyle=*}
\psdots(0.6,1.4)(1.5,1.4)
\psset{dotsize=3pt}\psset{dotstyle=o}
\psdots(0.9,1.7)(1.2,1.7)

%GRAPHIC 8b
\psline{-}(3.3,1.1)(3.6,1.4) \psline{-}(3.6,1.4)(3.9,1.7)
\psline{-}(3.9,1.7)(4.2,1.7) \psline{-}(4.2,1.7)(4.5,1.4)
\psline{-}(4.5,1.4)(4.8,1.7) \psline{-}(4.8,1.7)(5.1,1.4)
\psline{-}(5.1,1.4)(5.4,1.1)

%dots
\psset{dotsize=3pt}\psset{dotstyle=*}
\psdots(3.6,1.4)(5.1,1.4)
\psset{dotsize=3pt}\psset{dotstyle=o}
\psdots(3.9,1.7)(4.2,1.7)

%GRAPHIC 8c
\psline{-}(6.3,1.1)(6.6,1.4) \psline{-}(6.6,1.4)(6.9,1.7)
\psline{-}(6.9,1.7)(7.2,1.4) \psline{-}(7.2,1.4)(7.5,1.7)
\psline{-}(7.5,1.7)(7.8,1.7) \psline{-}(7.8,1.7)(8.1,1.4)
\psline{-}(8.1,1.4)(8.4,1.1)

%dots
\psset{dotsize=3pt}\psset{dotstyle=*}
\psdots(6.6,1.4)(8.1,1.4)
\psset{dotsize=3pt}\psset{dotstyle=o}
\psdots(7.5,1.7)(7.8,1.7)

%GRAPHIC 8d
\psline{-}(9.3,1.1)(9.6,1.4) \psline{-}(9.6,1.4)(9.9,1.1)
\psline{-}(9.9,1.1)(10.2,1.4) \psline{-}(10.2,1.4)(10.5,1.7)
\psline{-}(10.5,1.7)(10.8,1.7) \psline{-}(10.8,1.7)(11.1,1.4)
\psline{-}(11.1,1.4)(11.4,1.1)

%dots
\psset{dotsize=3pt}\psset{dotstyle=*}
\psdots(10.2,1.4)(11.1,1.4)
\psset{dotsize=3pt}\psset{dotstyle=o}
\psdots(10.5,1.7)(10.8,1.7)

\end{pspicture}
\end{center}
\end{figure}
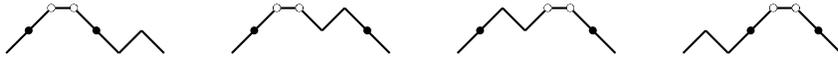

\subsection{Mixing the particles of charge 1/2 within the minimal-weight configuration}

The particles of charge 1/2 and 1 do not mix together, that is,  they cannot interpenetrate. 
Indeed, an H edge cannot be inserted within a charge 1 particle since that would transform it into a particle of charge 3/2. Phrased differently, the minimal distance between them must be 1, which prevents any interpenetration.
%YY
The combinatorial analysis of all possible insertions of these peaks of charge 1/2 and 1  within the higher charged particles can thus be made independently.

The insertion of the $n_{1/2}$ particles of charge 1/2 within those of charge $\geq 3/2$, taking into account the weight increase, is given by the $q$-binomial factor
\begin{equation}\label{demi}
%\begin{bmatrix} n_{\frac12}-1+\sum_{j=3/2}^{k/2+1} 2(j-1)n_{j}  \\ n_{\frac{1}2}\end{bmatrix}_q\equiv  
\begin{bmatrix}
n_{\frac12}-1+M \\ n_{\frac{1}2}\end{bmatrix}_q ,
\end{equation}
with $M$ defined in (\ref{m1}).
In dealing with such a $q$-binomial, we would need to impose also the requirement that $\begin{bmatrix}
-1\\ 0\end{bmatrix}_q =1$.

%Here is the justification.

The above result  can be justified as follows. 
The number of possible distinct insertions of an H edge within a particle of charge $ j>1$ is easily seen to be $2(j-1)$. This holds for both parities of $2j$. For $j$ integer, the H edge can be placed within any pair of (NE,NE) or (SE,SE) edges and there are $2(j-1)$ such vertices.
However,
 the H edge cannot be located on the top, at the height $j$,
 %(i.e., in-between the NE and SE edges closing the peak)
 since that would modify
 the resulting composite into a particle of charge $j+1/2$. Phrased differently, this is ruled out by the minimal distance condition -- which is 1 in this case --, since a top H edge  would give a separation of 1/2. 

If $j$ is half-integer, an H edge can be  inserted between any pair of (NE,NE) or (SE,SE)  edges; there are $2(\lf j\rf -1)=2j-2$ of theses. But in addition, an H edge can be inserted on  the top (beside the H edge already present)  since the  resulting separation between the two particles, which is the separation between the middle of the two H edges, is then 1, which is allowed.  There are actually two insertion points for this top edge but they result into identical configurations and should not be counted twice.

More generally, the insertion of  one H edge within $n_j$ particles of charge $j$ gives $2(j-1)n_j$ possibilities. When the number of H edges is  $n_{1/2}$,  this is generalized to 
\begin{equation}
\begin{pmatrix} n_{1/2}-1+2(j-1)n_j\\ n_{1/2}\end{pmatrix}.
\end{equation}  
That  the $n_{1/2}$  H edges representing the particles of charge 1/2  must  be placed inside one of the particle of type $j$ from the start, is the source of the factor $-1$.  Note also that the H edges cannot be placed in-between two particles: they must  be inserted within a specific particle. This combinatorial factor is illustrated in Fig. \ref{fig8} with $n_{1/2}=n_3=1$.

A straightforward generalization of this counting argument for the insertion of H edges within a set of particles of different charges accounts for the  combinatorial factor 
\begin{equation}
\begin{pmatrix} n_{1/2}-1+M\\ n_{1/2}\end{pmatrix}.
\end{equation}  
which is  (\ref{demi}) when $q=1$. 

To verify the correctness of the $q$-deformation of this binomial factor, we note that each displacement of an H edge inside a particle, from right to left, increases the weight by 1. Indeed, the effect of moving the inserted H edge toward the left within the particle is equivalent  to displace the position of a vertex of type (SE,SE) or (NE,NE) by two units toward the right while displacing two corners (which contribute half their position) by two units toward the left , so that $\Delta w= 2-2(1/2)=1$. This is also true when the H edge located at the first (NE,NE) vertex of a particle is moved to the last (SE,SE) vertex of the adjacent particle at its left.
% (see for instance Fig. \ref{fig8}).

\begin{figure}[ht]
\caption{{\footnotesize The various displacements of  a particle of type $1/2$ through a particle of type 3. At each step, a black dot is moved toward the right by 2 units and two circled dots are displaced by one unit each toward the left, for a weight difference of 1. }} \label{fig8}
\begin{center}
\begin{pspicture}(0,0)(12.0,2.6)

%GRAPHIC 7a
\psline{-}(0.3,1.1)(0.6,1.4) \psline{-}(0.6,1.4)(0.9,1.7)
\psline{-}(0.9,1.7)(1.2,2.0) \psline{-}(1.2,2.0)(1.5,1.7)
\psline{-}(1.5,1.7)(1.8,1.4) \psline{-}(1.8,1.4)(2.1,1.4)
\psline{-}(2.1,1.4)(2.4,1.1)

%dots
\psset{dotsize=3pt}\psset{dotstyle=*}
\psdots(0.6,1.4)(0.9,1.7)(1.5,1.7)
\psset{dotsize=3pt}\psset{dotstyle=o}
\psdots(1.8,1.4)(2.1,1.4)

%GRAPHIC 7b
\psline{-}(3.3,1.1)(3.6,1.4) \psline{-}(3.6,1.4)(3.9,1.7)
\psline{-}(3.9,1.7)(4.2,2.0) \psline{-}(4.2,2.0)(4.5,1.7)
\psline{-}(4.5,1.7)(4.8,1.7) \psline{-}(4.8,1.7)(5.1,1.4)
\psline{-}(5.1,1.4)(5.4,1.1) 

%dots
\psset{dotsize=3pt}\psset{dotstyle=*}
\psdots(3.6,1.4)(3.9,1.7)(5.1,1.4)
\psset{dotsize=3pt}\psset{dotstyle=o}
\psdots(4.5,1.7)(4.8,1.7)

%GRAPHIC 7c
\psline{-}(6.3,1.1)(6.6,1.4) \psline{-}(6.6,1.4)(6.9,1.7)
\psline{-}(6.9,1.7)(7.2,1.7) \psline{-}(7.2,1.7)(7.5,2.0)
\psline{-}(7.5,2.0)(7.8,1.7) \psline{-}(7.8,1.7)(8.1,1.4)
\psline{-}(8.1,1.4)(8.4,1.1)

%dots
\psset{dotsize=3pt}\psset{dotstyle=*}
\psdots(6.6,1.4)(7.8,1.7)(8.1,1.4)
\psset{dotsize=3pt}\psset{dotstyle=o}
\psdots(6.9,1.7)(7.2,1.7)

%GRAPHIC 7d
\psline{-}(9.3,1.1)(9.6,1.4) \psline{-}(9.6,1.4)(9.9,1.4)
\psline{-}(9.9,1.4)(10.2,1.7) \psline{-}(10.2,1.7)(10.5,2.0)
\psline{-}(10.5,2.0)(10.8,1.7) \psline{-}(10.8,1.7)(11.1,1.4)
\psline{-}(11.1,1.4)(11.4,1.1) 

%dots
\psset{dotsize=3pt}\psset{dotstyle=*}
\psdots(10.2,1.7)(10.8,1.7)(11.1,1.4)
\psset{dotsize=3pt}\psset{dotstyle=o}
\psdots(9.6,1.4)(9.9,1.4)

\end{pspicture}
\end{center}
\end{figure}
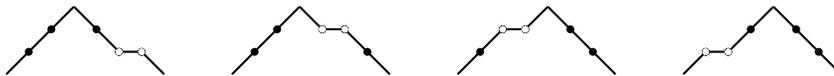

 Let us indicate a compatibility test of our procedure: The insertion of the 1/2 particles within particles of charge $\geq 3/2$  can be done before or after these have been mixed and the equivalence of the different countings relies crucially upon the fact that H edges representing the charge 1/2 particles cannot be inserted in-between the particles.
  It will suffice to illustrate this with an example. Consider the insertion of a particle 1/2 into different arrangements of two particles of charge 3 and 3/2. 
  % The number of allowed insertions should be independent of the way the particles are interpenetrated. 
  Three typical configurations are displayed in Fig. \ref{fig9}. The insertion points where the charge 1/2 particle can be placed are indicated by crossed circles. We see that their number is the same for each configuration. This would clearly not be so  if we had allowed insertions of H edges in-between the particles. In the second configuration, that would lead to an extra possibility. In the third one, the in-between insertion of an H edge does not produce a distinct configuration and it would thus not be counted. Moreover,  when the particles are separated 
  as in the first configuration,  an in-between H edge would simply not be allowed by the path admissibility conditions (\ref{adm}) since it  would then lie at the lower boundary of the defining strip.  The counting would thus depend upon the relative position of the two particles.

\begin{figure}[ht]
\caption{{\footnotesize The possible insertion points, indicated by crossed circles, for a particle of charge $1/2$ in three  different configurations involving two particles of respective charge $3$ and $3/2$. The number of possible insertions  is independent of the interpenetration pattern of the two higher charge particles.}} \label{fig9}
\begin{center}
\begin{pspicture}(0,0)(9.0,3.0)

%GRAPHIC 9a
\psline{-}(0.3,1.1)(0.6,1.4) \psline{-}(0.6,1.4)(0.9,1.7)
\psline{-}(0.9,1.7)(1.2,2.0) \psline{-}(1.2,2.0)(1.5,1.7)
\psline{-}(1.5,1.7)(1.8,1.4) \psline{-}(1.8,1.4)(2.1,1.1)
\psline{-}(2.1,1.1)(2.4,1.4) \psline{-}(2.4,1.4)(2.7,1.4) 
\psline{-}(2.7,1.4)(3.0,1.1) 

%dots
\psset{dotsize=4pt}\psset{dotstyle=o}
\psdots(0.6,1.4)(0.9,1.7)(1.5,1.7)(1.8,1.4)(2.7,1.4)
\psset{dotstyle=+}\psset{dotangle=45} 
\psdots(0.6,1.4)(0.9,1.7)(1.5,1.7)(1.8,1.4)(2.7,1.4)

%GRAPHIC 9b
\psline{-}(3.3,1.1)(3.6,1.4) \psline{-}(3.6,1.4)(3.9,1.7)
\psline{-}(3.9,1.7)(4.2,2.0) \psline{-}(4.2,2.0)(4.5,1.7)
\psline{-}(4.5,1.7)(4.8,1.4) \psline{-}(4.8,1.4)(5.1,1.7)
\psline{-}(5.1,1.7)(5.4,1.7) \psline{-}(5.4,1.7)(5.7,1.4) 
\psline{-}(5.7,1.4)(6.0,1.1) 

%dots
\psset{dotsize=4pt}\psset{dotstyle=o}
\psdots(3.6,1.4)(3.9,1.7)(4.5,1.7)(5.4,1.7)(5.7,1.4)
\psset{dotstyle=+}\psset{dotangle=45} 
\psdots(3.6,1.4)(3.9,1.7)(4.5,1.7)(5.4,1.7)(5.7,1.4)

%GRAPHIC 9c
\psline{-}(6.3,1.1)(6.6,1.4) \psline{-}(6.6,1.4)(6.9,1.7)
\psline{-}(6.9,1.7)(7.2,2.0) \psline{-}(7.2,2.0)(7.5,1.7)
\psline{-}(7.5,1.7)(7.8,1.7) \psline{-}(7.8,1.7)(8.1,2.0)
\psline{-}(8.1,2.0)(8.4,1.7) \psline{-}(8.4,1.7)(8.7,1.4)
\psline{-}(8.7,1.4)(9.0,1.1)

%dots
\psset{dotsize=4pt}\psset{dotstyle=o}
\psdots(6.6,1.4)(6.9,1.7)(7.8,1.7)(8.4,1.7)(8.7,1.4)
\psset{dotstyle=+}\psset{dotangle=45} 
\psdots(6.6,1.4)(6.9,1.7)(7.8,1.7)(8.4,1.7)(8.7,1.4)

\end{pspicture}
\end{center}
\end{figure}
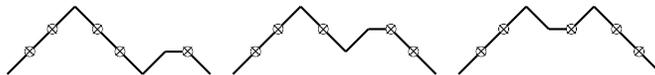

  Within the framework of a fermi-gas description of the path (that is, irrespective of the operator formulation to be spelled out later), this is basically the argument legitimating the rule for the distribution of the particles of charge 1/2 within mixtures of higher charge particles. This in turn justifies the way we have allowed  the particles to be slightly deformed in the penetration process: this ensures that all possible configurations are reached.\footnote{Note that with our rules, the insertion of the H edge in-between the two particles in the second configuration of Fig. \ref{fig9} amounts to change the charge of the particle, from  3/2 to 2. This configuration is effectively  taken into account but when summing over the charge content.}

%and for the special way

    %with $p_3=p_{3/2}=1$ of a particle of charge

\subsection{Finitized vacuum character}

All configurations can be generated from the minimal-weight configurations by mixing the particles the way it has been described in the previous three subsections. The generating function for all the paths
with $(\ell_0,\ell_L)=(1,1)$ with the different $n_j$ fixed
%, and such that the cumulative diameter of all particles add up to $L$,
%(and recall that the length is $L=2\sum_i  i n_i$) 
is thus given by the product of all these combinatorial factors
\begin{equation}
q^{w_{{\rm mwc}}}\begin{bmatrix}
n_{\frac12}-1+M \\ n_{\frac{1}2}\end{bmatrix}_q 
\begin{bmatrix}
n_1+M   \\ n_1 \end{bmatrix}_{q^2} \prod_{i=3/2}^{k/2+1}
\begin{bmatrix}
n_{i}+m_i  \\ n_{i}\end{bmatrix}_q.
%\qquad (\text{with}\quad \sum_{i=1/2}^{k/2+1}2 in_i = L).
\end{equation}
The generating function for all paths of length $L$ is then obtained by summing over all  the numbers $n_j$ that are compatible with the fixed length condition and this gives
\begin{equation}\label{vacuf}
%\chi_{1,1}^{(L)}(q)= 
\sum_{\substack{n_{\frac12}, n_{1},\cdots , n_{\frac{k}2+1} =0\\n_{\frac12}+2n_1+\cdots+(k+2)n_{\frac{k}2+1} = L}}^\y
q^{ nBn +\frac12n_{\frac12}}\begin{bmatrix}
n_{\frac12}-1+M \\ n_{\frac{1}2}\end{bmatrix}_q 
\begin{bmatrix}
n_1+M   \\ n_1 \end{bmatrix}_{q^2} \prod_{i=3/2}^{k/2+1}
\begin{bmatrix}
n_{i}+m_i  \\ n_{i}\end{bmatrix}_q, 
\end{equation}
with $B$, $M$, and $m_i$ given respectively in (\ref{Bij}), (\ref{m1}) and (\ref{mi}).  For $L$ even, this generates only integer powers of $q$. Since the vacuum is in the NS sector, there are also states at half-integer conformal dimension. These are recovered by taking  $L$ odd. Therefore,  to obtain the full finitized version of the vacuum character, one needs to add the contribution of $L$ and $L+1$.
% as the for $L$ even, we get the 
% integer powers of $q$ and for $L$ odd, the half-integer powers. 

%\subsection{An identity}

\subsection{Character formulae in the infinite length limit}

Consider the infinite length limit obtained by setting the sum $n_{1/2}+2 n_1\rw \y$ or equivalently, by setting $P=n_{1/2}+ n_1\rw \y$. Note that the variables $n_{1/2}$ and $n_1$ enter in the above expression only in the first two $q$-binomial factors. Taking the infinite length limit amounts then to evaluate the sum (with $n_{1/2}=n$ and $n_1=P-n$)
\begin{equation}
S= \lim_{P\rw \y} \sum_{n=0}^P q^{n/2}
\begin{bmatrix}
n- 1+M  \\ n \end{bmatrix}_q 
\begin{bmatrix}
P-n+M  \\ M \end{bmatrix}_{q^2},
\end{equation}
where 
$M$ is defined in (\ref{m1}).
Thanks to the converging prefactor $q^{n/2}$ (with the tacit assumption that $|q|<1$), the limit can be taken directly and it yields:
\begin{equation}
S= \frac{1}{(q^2;q^2)_{M} }\sum_{n=0}^\y q^{n/2}
\begin{bmatrix}
n- 1+M  \\ n \end{bmatrix}_q ,\end{equation}
(recall the definition (\ref{zqde})).
The summation is then recognized as the expansion of the inverse of the $q$-factorial $(q^{1/2})_M$ (cf. Theorem 3.3 in \cite{Andr}):
\begin{equation}\label{Siden}
S= \frac{1}{(q^2;q^2)_{M} (q^{1/2})_M}= \frac{(-q^{1/2})_M\,  (q^{1/2})_M}{(q)_{2M} \, (q^{1/2})_M} =  \frac{(-q^{1/2})_M}{(q)_{2M}}.
\end{equation}

By redefining the summation variables as $n_{i/2}=p_{i-2}$, so that the multiple-sum indices are now 
$p_1,\cdots , p_k$, with  $M= \sum_{i=1}^k i p_i $, we have
\begin{equation}\label{us}
\chi_{1,1}(q)= \sum_{p_1,\cdots, p_k\geq 0}\frac{q^{pB'p}(-q^{1/2})_M} {(q)_{2M} }
\prod_{i=1}^{k-1}
\begin{bmatrix}
p_i+\sum_{j=1}^{k-i} 2jp_{j+i}  \\ p_{i}\end{bmatrix}_q, 
\end{equation}
where $B'$ is the symmetric matrix with entries $B_{ij}'= i(j+2)/2$ for $i\leq j$. 
This is the vacuum character of the superconformal minimal model $ \SM(k+2,k+4)$.

For the first three models $(k=1,2,3)$, that is, for the $\mathcal {SM}(3,5)$, $\mathcal {SM}(4,6)$ and the $\mathcal {SM}(5,7)$ model respectively,  the explicit form of the vacuum character reads:
\begin{align} \label{k123}
&  \sum_{p_1=0}^\y   \frac{ q^{\frac32 p_1^2} (-q^{1/2})_{p_1} }{(q)_{2p_1} },
\nonumber\\
& \sum_{p_1,p_2=0}^\y  \frac{ q^{\frac32 p_1^2+4p_2^2+4p_1p_2}  (-q^{1/2})_{p_1+2p_2} }{(q)_{2p_1+4p_2} }
 \begin{bmatrix}
p_1+ 2p_2\\ p_1\end{bmatrix}_q ,
\nonumber\\
& \sum_{p_1,p_2,p_3 =0}^\y \frac{
q^{ \frac32p_1^2+4p_2^2+\frac{15}2p_3^2+4p_1p_2+5p_1p_3+10p_2p_3}(-q^{1/2})_{p_1+2p_2+3p_3}} { (q)_{2p_1+4p_2+6p_3} } 
\begin{bmatrix}
p_1+2p_2+4p_3\\ p_1\end{bmatrix}_q
\begin{bmatrix}
p_2+ 2p_3\\ p_2\end{bmatrix}_q.
\end{align}

\subsection{Comparison with known fermionic formulae}

The vacuum character of the $\SM(k+2,k+4)$ given in \cite{BG,Sch} takes the form
\begin{equation}\label{gep}
\chi_{1,1}= \sum_{\substack{\ell_1,\cdots, \ell_{k+1} \geq 0\\ \ell_i  \,\text{ even for $i\geq 2$}}}\frac{q^{\frac14 \ell C \ell }} {(q)_{\ell_2} }
\prod_{\substack{i=1\\i\not=2}}^{k+1}
\begin{bmatrix}
\frac12(\ell_{i-1}+\ell_{i+1})  \\ \ell_{i}\end{bmatrix}_q, 
\end{equation}
(with the understanding that  $\ell_0=\ell_{k+2}=0$). Here $C$ is the $A_{k+1}$ Cartan matrix ($C_{i,i}=2, \,C_{i-1,i}=C_{i,i+1}=-1$) and
 \begin{equation}
\ell C \ell = \sum_{i,j=1}^{k+1} \ell_i C_{ij} \ell_j .
\end{equation}
 To relate (\ref{us}) and (\ref{gep}), we first use the identity
\begin{equation}
(-q^{1/2})_M= \sum_{j=0}^M q^{j^2/2}
\begin{bmatrix}
M  \\j \end{bmatrix}_q 
\end{equation}
Then the comparison of the  two $q$-factorials in the denominator and the
products of $q$-binomials lead to the following relation between the two sets of variables $\{\ell_i\}$ and $\{j,\, p_i\}$:
\begin{equation}
\ell_1= M-j, \quad \ell_2 = 2M \quad \text{and}\quad  \ell_i= 2\sum_{j=0}^{k-i+1} (j+1)p_{j+i-1} \quad \text{for $i\geq 3$}. 
\end{equation}
(For instance, for $k=4$, $\ell_3=2p_2+4p_3$ and $\ell_4=2p_3$.) Note that the variables $\ell_i$ so defined are such  that $\ell_1$ is a non-negative integer while $\ell_2,\cdots \ell_{k+1}$ are all even. It is then straightforward to verify the equality of the $q$-exponent in the numerators:
\begin{equation}
 \frac14( \ell C \ell -2 j^2) = pB'p.
 \end{equation} 
 Note that this change of variables involves a summation mode $j$ that has no interpretation in terms of the path.
This hints that there is no direct relation between the finitized versions of \cite{BG,Sch} and ours. 
%Actually, there is no expression relating $L$ and the $\ell_i$. 
It appears thus that the formula given in (\ref{vacuf}) is new.\footnote{
In order to see the structural difference between the two finitized expressions in the simplest context, 
consider the expression for the finite version of the vacuum $\SM(3,5)$ character. In \cite{BG,Sch}, it reads:
\begin{equation*}\label{gep35}
{\tilde\chi}_{1,1}^{(3,5)}(q;L)= \sum_{\substack{\ell_1,\ell_{2} \geq 0\\L+\ell_1,\, \ell_2  \,\text{ even} }}\, q^{\frac12( \ell_1^2+\ell_2^2- \ell_1\ell_2) }\begin{bmatrix}
\frac12\ell_{2}  \\ \ell_{1}\end{bmatrix}_q
\begin{bmatrix}
\frac12(\ell_{1}+L)  \\ \ell_{2}\end{bmatrix}_q.
\end{equation*}
This is to be compared with 
\begin{equation*}
{\tilde\chi}_{1,1}^{(3,5)}(q;L)= \sum_{\substack{n_{\frac12}, n_{1}, n_{\frac{3}2} =0\\n_{\frac12}+2n_1+3n_{\frac{3}2} = L}}^\y
q^{\frac32 n_{\frac32}^2+\frac12 n_{\frac12} }\begin{bmatrix}
n_{\frac12}-1+n_{\frac32}\\ n_{\frac{1}2}\end{bmatrix}_q 
\begin{bmatrix}
n_1+n_{\frac32}   \\ n_1 \end{bmatrix}_{q^2} .%\qquad (\text{with}\quad \sum_{i=1/2}^{k/2+1}2 in_i = L).
\end{equation*}
In both cases, with $L$ even (odd), we recover integer (half-integer) powers of $q$, hence the tilde on the character (reminding that this is not the full finitized character). These formulae give identical expansions (for $L\leq 50$ and to order 100). Note that in \cite{BG,Sch} the denominator $(q)_{\ell_2}$ appearing in (\ref{gep})
 is the infinite length limit of the $q$-binomial
\begin{equation}
\frac{1}{ (q)_{\ell_2} }  = \lim_{L\rw \y}
\begin{bmatrix}
\frac12(\ell_{1}+\ell_{3}+L)  \\ \ell_{2}\end{bmatrix}_q. 
\end{equation}
This $q$-binomial factor captures the whole dependence of the finite form upon $L$.} Further support for this claim is provided in App. A where the infinite limit of the dual character (obtained by $q\to q^{-1}$) is shown to lead a new form of the parafermonic $\z_{k+2}$ characters. The difference with the usual expression \cite{LP} -- which  can be obtained by duality for the finite characters of \cite{BG,Sch} as demonstrated in \cite{Sch} --, indicates a difference between the two finite versions since in the dual case, the infinite length limit is taken differently, that is, via $n_{k/2+1}\to \y$. In particular, that our finite  character involves a $q^2$-binomial factor reflects itself in the corresponding parafermionic character by the presence of a non-usual $(q^2;q^2)$-factorial term.

%\section{Description of the other modules}

\section{Operator construction}

\subsection{Interpreting a path as a sequence of non-local operators}

In \cite{JMop}, we have derived an operator interpretation for the paths describing the unitary minimal models (cf. \cite{FP} for an earlier approach along these lines). The idea is to associate  to all contributing vertices, the action of an operator acting at the corresponding (horizontal) position. This position fixes the mode index of the operator. These operators are subject to certain rules that are readily lifted from the mere characteristics of the  path. These rules constitute the basis relations.

The same approach can be followed here. One introduces two types of operators, $a$ and $a^*$. The local part of their action is described as follows. Let us again indicate the  vertex in-between edges of type A and B as (A,B). Locally, $a$ acts on vertices of type (NE,SE)  or (NE,H), to transform them respectively into (NE,H) and (NE,NE), that is:
\begin{align}
a: \; &\text{(NE,SE)$\, \rw\, $ (NE,H)}\nonumber\\
 \; &\text{(NE,H)\;\,$\,\rw\, $ (NE,NE).}
 \end{align}
 We denote by $a_i$ the operator acting at $i$; it has weight $i/2$. Two operators can act at $i$ if it is the position of a maximum: $a^2_i$ transforms then (NE,SE) into (NE,NE) (and it has total weight $i$).  Note however that three operators or more cannot act at the same point: $a_i^n=0$ for $n>2$.

Similarly, the local component of the  $a^* $ action is as follows:
%on (SE,NE) to change it into (SE,H) and on (SE,H) to transform it into (SE,SE):
\begin{align}
a^*: \; &\text{(SE,NE)$\, \rw\, $ (SE,H)}\nonumber\\
 \; &\text{(SE,H)\;\;\,$\,\rw\, $ (SE,SE).}
 \end{align}
Again, its mode index indicates the position where it acts, which is twice its weight.

Like their non-supersymmetric counterparts \cite{JMop}, these operators actually  act in a very non-local way: their action affects  the path from the point of application to its right extremity.
The action of $a_i $ on the ground-state amounts to change all edges at the right of $i$ onto H edges, that is, it transforms the zig-zag path tail into a straight line. A further action of $a_i$ undo the flattening of  the path by reinserting the zig-zag pattern but lifted upward by one unit compared to its ground-state position.  The full action of $a_i^2$ on the ground-state is thus to create a NE edge at $i$, translate the tail of the path by one unit both vertically and horizontally and remove the last edge. This is akin to the action of the operator $b$ defined in \cite{JMop}.\footnote{
%Note however that the notation used in \cite{JMop} differs somewhat from the more convenient one chosen here (which avoids fractional modes): there the mode index is the weight and thus half the acting position: therefore 
The present 
 $a_i^2$ is similar (in its action on the ground state) to the operator $b_{i}$ in \cite{JMop}.}

The operator $a^*_i$ acts similarly but in the opposite vertical direction. 
 %Note that  the right boundary is generally modified by those actions. To preserve the boundary conditions of the ground-stae, a seqence 
Sample actions on the ground (or vacuum)  state are pictured in Fig. \ref{fig11}.

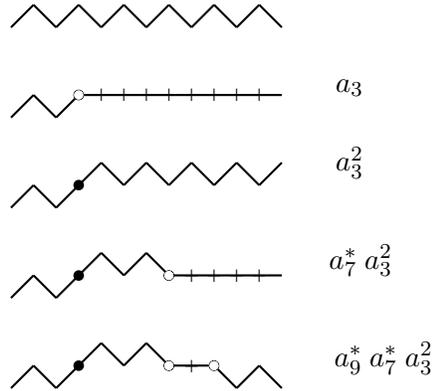
\begin{figure}[ht]
\caption{{\footnotesize The successive actions of the sequence of operators $a^*_9 a^*_7 a_3^2$ on the ground-state path displayed in the first line. The action of each operator is manifestly non-local, modifying the path from its insertion point to its right extremity.}} \label{fig11}
\begin{center}
\begin{pspicture}(0,0)(6.0,6.0)

%GRAPHIC 10a
\psline{-}(0.3,4.8)(0.6,5.1) \psline{-}(0.6,5.1)(0.9,4.8)
\psline{-}(0.9,4.8)(1.2,5.1) \psline{-}(1.2,5.1)(1.5,4.8)
\psline{-}(1.5,4.8)(1.8,5.1) \psline{-}(1.8,5.1)(2.1,4.8)
\psline{-}(2.1,4.8)(2.4,5.1) \psline{-}(2.4,5.1)(2.7,4.8) 
\psline{-}(2.7,4.8)(3.0,5.1) \psline{-}(3.0,5.1)(3.3,4.8) 
\psline{-}(3.3,4.8)(3.6,5.1) \psline{-}(3.6,5.1)(3.9,4.8) 

%dots

%GRAPHIC 10b
\psline{-}(0.3,3.6)(0.6,3.9) \psline{-}(0.6,3.9)(0.9,3.6)
\psline{-}(0.9,3.6)(1.2,3.9) \psline{-}(1.2,3.9)(1.5,3.9)
\psline{-}(1.5,3.9)(1.8,3.9) \psline{-}(1.8,3.9)(2.1,3.9)
\psline{-}(2.1,3.9)(2.4,3.9) \psline{-}(2.4,3.9)(2.7,3.9) 
\psline{-}(2.7,3.9)(3.0,3.9) \psline{-}(3.0,3.9)(3.3,3.9) 
\psline{-}(3.3,3.9)(3.6,3.9) \psline{-}(3.6,3.9)(3.9,3.9) 

%dots
\psset{dotsize=4pt}\psset{dotstyle=o}
\psdots(1.2,3.9)
\psset{dotsize=4pt}\psset{dotstyle=+}
\psdots(1.5,3.9)(1.8,3.9)(2.1,3.9)(2.4,3.9)(2.7,3.9)(3.0,3.9)(3.3,3.9)(3.6,3.9)

%GRAPHIC 10c
\psline{-}(0.3,2.4)(0.6,2.7) \psline{-}(0.6,2.7)(0.9,2.4)
\psline{-}(0.9,2.4)(1.2,2.7) \psline{-}(1.2,2.7)(1.5,3.0)
\psline{-}(1.5,3.0)(1.8,2.7) \psline{-}(1.8,2.7)(2.1,3.0)
\psline{-}(2.1,3.0)(2.4,2.7) \psline{-}(2.4,2.7)(2.7,3.0) 
\psline{-}(2.7,3.0)(3.0,2.7) \psline{-}(3.0,2.7)(3.3,3.0) 
\psline{-}(3.3,3.0)(3.6,2.7) \psline{-}(3.6,2.7)(3.9,3.0) 

%dots
\psset{dotsize=4pt}\psset{dotstyle=*}
\psdots(1.2,2.7)

%GRAPHIC 10d
\psline{-}(0.3,1.2)(0.6,1.5) \psline{-}(0.6,1.5)(0.9,1.2)
\psline{-}(0.9,1.2)(1.2,1.5) \psline{-}(1.2,1.5)(1.5,1.8)
\psline{-}(1.5,1.8)(1.8,1.5) \psline{-}(1.8,1.5)(2.1,1.8)
\psline{-}(2.1,1.8)(2.4,1.5) \psline{-}(2.4,1.5)(2.7,1.5) 
\psline{-}(2.7,1.5)(3.0,1.5) \psline{-}(3.0,1.5)(3.3,1.5) 
\psline{-}(3.3,1.5)(3.6,1.5) \psline{-}(3.6,1.5)(3.9,1.5) 

%dots
\psset{dotsize=4pt}\psset{dotstyle=*}
\psdots(1.2,1.5)
\psset{dotsize=4pt}\psset{dotstyle=o}
\psdots(2.4,1.5)
\psset{dotsize=4pt}\psset{dotstyle=+}
\psdots(2.7,1.5)(3.0,1.5)(3.3,1.5)(3.6,1.5)

%GRAPHIC 10e
\psline{-}(0.3,0.0)(0.6,0.3) \psline{-}(0.6,0.3)(0.9,0.0)
\psline{-}(0.9,0.0)(1.2,0.3) \psline{-}(1.2,0.3)(1.5,0.6)
\psline{-}(1.5,0.6)(1.8,0.3) \psline{-}(1.8,0.3)(2.1,0.6)
\psline{-}(2.1,0.6)(2.4,0.3) \psline{-}(2.4,0.3)(2.7,0.3) 
\psline{-}(2.7,0.3)(3.0,0.3) \psline{-}(3.0,0.3)(3.3,0.0) 
\psline{-}(3.3,0.0)(3.6,0.3) \psline{-}(3.6,0.3)(3.9,0.0) 

%dots
\psset{dotsize=4pt}\psset{dotstyle=*}
\psdots(1.2,0.3)
\psset{dotsize=4pt}\psset{dotstyle=o}
\psdots(2.4,0.3)(3.0,0.3)
\psset{dotsize=4pt}\psset{dotstyle=+}
\psdots(2.7,0.3)

%text
\rput(4.8,4.0){{\small $a_3$}} 
\rput(4.8,3.0){{\small $a^2_3$}}
\rput(5.0,1.7){{\small $a^{*}_7$ $a^2_3$ }}
\rput(5.3,0.4){{\small  $a^{*}_9$ $a^{*}_7$ $a^2_3$ }}

\end{pspicture}
\end{center}
\end{figure}

Any path can thus be represented by a string of operators acting on the vacuum ground-state path. It will be understood that these are ordered in increasing values of their modes from left to right. Observe that in order to preserve the boundary conditions of the ground-state path,
%for the path to start from the point $(0,1)$ and end on the point $(L,1)$, 
the string must contain  an equal  number of $a$ and $a^*$ operators.  For instance, the path of Fig. 3 is associated to the following sequence of operators: $a_{26}^{*2}\cdots a_9^{*2} a_7 a^*_5 a^*_4  a_3 a_2a_1$.

As already indicated, the path characteristics capture all the conditions that need to be imposed on sequences of operators in order to define a basis. An analysis of this type has already been presented in \cite{JMop} and it can be adapted  to the present context with minor modifications. Instead of following this line of presentation, we will proceed in a more informal way by treating the first few models explicitly. From these considerations, the general pattern will emerge naturally. Note also that  the operator method, taken independently of its path origin, is not naturally finitized. In a first step, we thus avoid this finitization constraint and build up directly the $L\rw \y$ expression of the characters. Our main objective is not much to be complete and systematic  concerning the operator construction but rather to indicate how it makes natural our previous particle identification within a path.

\subsection{The operator description of the $\SM(3,5)$ model}

For $k=1$, the first excitation in the vacuum module corresponds to a peak of charge 3/2 centered at the position 3/2. This value of the charge is also the maximal allowed height for a peak in this model. It is understood that the path is infinite, being completed by an infinite sequence of peaks of charge 1. In our operator terminology, this peak of charge 3/2 is equivalent to the string of operators $a^*_2a_1$ (of weight 3/2,  half the sum of the indices), acting on the ground state. The various states that can be reached from this elementary sequence are simply those obtained by increasing the modes (or displacing them, using a pictorial terminology that carries a path flavor) in all possible ways by respecting their ordering. The order needs to be preserved since, from the definition of its action,  $a^*$ cannot act first, that is, directly on the vacuum state. This fixed-ordering  condition is interpreted as a hard-core repulsion between the two modes.

Let us describe explicitly  the different  configurations that can be obtained from $a^*_2a_1$ and derive their relative weight. The $a^*$ mode can be increased by any integer or displaced by any number of steps and each unit displacement increases the weight by 1/2. In the path terminology, a move of $a^*$ by $m$ steps  amounts to insert $m$ particles of charge 1/2 (or $m$ H edges) within the particle of charge $3/2$. 
%For instance, an increase by 2 yields to the string $a_1a^*_4$.  
The generating function for all the displacements of the $a^*$ mode is $(1-q^{1/2})^{-1}$. The $a$ mode can also be increased but this must be twined with a similar augmentation of the $a^*$ mode, i.e., in its  
 displacement, $a$ drags the   $a^*$ mode. Moreover, the path interpretation of the move shows that is must be done by steps of two units: the path cannot start with an H edge, so a full  particle of charge 1 has to be inserted in front of the particle of charge 3/2. The generating function for these displacements is thus $(1-q^2)^{-1}$. Taking care of both type of displacements, as well of the weight of the lowest-weight configuration, the generating function for all states described by a single pair of $a^*a$ operators is
 \begin{equation}  
\frac{q^{\frac32 }}{  (1-q^{\frac12}) (1-q^{2}) } .
\end{equation}

In order to see how this gets generalized to the case 
 where there are  
$n$ pairs of $a^*a$ operators,
consider the case where $n=2$. Again, the modes increase must preserve the operator ordering since an interchange would either lead to a $a^*$ acting on the vacuum or generate a peak higher than 3/2. The displacements of the last two modes are described as before. The leftmost $a^*$ mode can be moved by any integer and each unit displacement drags the  pair of $a^*a$ that acts after by one unit each so that the total weight increase is $3/2$. The generating function for displacements of this type is thus $(1-q^{3/2})^{-1}$. Finally, the first $a$ mode can be moved by steps of two units, with a weight change of 4 each times (since its motion drags the subsequent three modes) and these are generated by $(1-q^{4})^{-1}$. The lowest-weight string is $a^*_5a_4 a_2^*a_1$, with weight $6$. The generating function for all states with two pairs of $a^*a$ is thus:
 \begin{equation}  
\frac{q^{6} }{  (1-q^{1/2})(1-q^2)(1-q^{3/2})(1-q^4)} .
\end{equation}
 Proceeding along this way, it is rather immediate to see that the generalization to $n$ pairs is 
\begin{equation}q^{\frac32 n^2}  \prod_{i=1}^{n}
\frac{1}{  (1-q^{i-\frac12}) (1-q^{2i}) } .
\end{equation}
Using the identity
\begin{equation}
  \left(\prod_{i=1}^{n}
\frac{1}{  (1-q^{i-\frac12}) (1-q^{2i}) } \right) = \frac{(-q^\frac12)_n} { (q)_{2n}},
\end{equation}
this reduces to 
\begin{equation}q^{\frac32 n^2} \frac{(-q^\frac12)_n} { (q)_{2n}}.
\end{equation}
The full generating function, which is nothing but the vacuum character of the $\SM(3,5)$
 model, is obtained by summing over $n$:
\begin{equation}\chi_{1,1}^{(3,5)}(q)= \sum_{n=0}^\y  \frac{q^{\frac32 n^2}(-q^\frac12)_n} { (q)_{2n}}.
\end{equation}
This matches the first expression in (\ref{k123}), with $n=p_1$.

\subsection{The operator description of the $\SM(4,6)$ model}

For the case where $k=2$, the are two types of contributing peaks, those with charge 3/2 and 2.
In terms of operators, these correspond to the two basic structures: $a^*a$ and $a^{*2}a^2$.  These are called 1- and 2-blocks respectively. Denote by $p_1$ and $p_2$ their respective number.
The sequence of lowest-weight is the one with modes as closely packed as possible and in decreasing value of the block content.
%ordered as follows: $a^2a^{*2}\cdots a^2a^{*2} aa^*\cdots aa^*$.

Let us consider first the analysis of the elementary 2-block $a_3^{*2}a_1^2$ (of weight 4)  and the various displacements of its constituents. One $a^*$ can be moved by any units, with a weight increase given by half its displacement. The generating factor is $(1-q^{1/2})^{-1}$. Similarly, the two $a^*$ modes can be moved together but by two units each step (which amounts to insert each time a peak of charge 1 in between the $a^{*2}$ and $a^{2}$ modes). The weight change is 2 each step, leading to the generating  factor  $(1-q^{2})^{-1}$. Next, a $a$ mode can be displaced by steps of one unit, with each displacement dragging the two $a^*$operators: the weight difference is 3/2 for each step and the generating factor is $(1-q^{3/2})^{-1}$. Finally, the displacements of the two $a$ modes are generated by $(1-q^{4})^{-1}$.  The generating function of all states obtained form one 2-block is 
\begin{equation}
\frac{ q^4 }{(1-q^{\frac12})(1-q^{\frac32})(1-q^{2})(1-q^{4})} .
\end{equation}
The extension to the case of $p_2$ 2-blocks is rather immediate:
\begin{equation}q^{4p_2^2}  \prod_{i=1}^{2p_2}
\frac{1}{  (1-q^{i-\frac12}) (1-q^{2i}) } =  \frac{ q^{4p_2^2}(-q^\frac12)_{2p_2}} { (q)_{4{p_2}}}.
\end{equation}

Let us then  turn to the case where both types of blocks are present.
By considering first  the case $p_2=p_1=1$, one can easily figure out the weight generating factor that corresponds to the various displacements that maintain the operator order fixed and it is: 
\begin{equation}
\frac{ q^\frac{19}{2} }{(1-q^{\frac12})(1-q^{\frac32})(1-q^{\frac52})(1-q^{2})(1-q^{4})(1-q^{6})} .
\end{equation}
For a generic number of 1- and 2-blocks, the minimal-weight configuration is readily found to be $4p_2^2+4p_2 p_1 + 3p_1^2/2$. The generating factor for the order-preserving displacements  is:
\begin{equation}\label{coma}
  \frac{ q^{4p_2^2+4p_2 p_1 + \frac32p_1^2}(-q^\frac12)_{2p_2+p_1}} { (q)_{4{p_2+2p_1}}}.
\end{equation}

But this is not quite the complete result: although the ordering of the terms within each block is immutable,  it remains
to take into account the mixing of blocks, that is, the immersion of the smaller  block into the larger one. 
%Such mixing is not ruled out by the hard-core repulsion since this  repulsion is effective between charged operators. In the block-mixing operation,  one does not interchange the position of two charged operators but rather `commutes' within a block a charge-less combination, that is, moves within the larger block, the whole smaller block.
We argue, in the following subsection, that this gives rises to the multiplicative factor
\begin{equation}\label{comb}
\begin{bmatrix} p_1 + 2p_2\\p_1\end{bmatrix}.
\end{equation} 
The product of (\ref{coma}) and (\ref{comb}) summed over all values of $p_1$ and $p_2$ gives the full character. The resulting expression  is
found to be identical with the second expression in (\ref{k123}).

Granting the result (\ref{comb}), we see from these two examples, that the operator construction provides a rather efficient way of generating the fermionic form of the characters.

\subsection{Block mixing}

% must preserve the block identities.  
It is convenient to consider the block-mixing operation  in general terms, by considering the mixing of a $i$-block within a $j$-block, with $j>i$. A $j$-block is of the form $ a^{*j}a^j$ and its charge is $j/2+1$.
In the same way as the particle interpenetration imposes a minimal separation between the peaks, the immersed block  cannot be too close to the center of the larger block.\footnote{The following discussion parallels the one already presented for  paths. However, it is spelled out again in this novel context since its naturalness  is our original  rationale for the special particle deformation previously introduced in the interpenetration process.}
  This prevents the production of  higher-type blocks in the immersion process. Equivalently, it ensures that  the blocks preserve their identity.  The center of a  $j$-block is in-between the $a^*$ and $a$ modes.  Assign charge 1 to $a$ and $-1$ to $a^*$: $q(a)=-q(a^*)=1$.  Introduce the $\ell$-th partial charge of a sequence of operators $c_{i_m}\cdots c_{i_1}$, where $c$ is either $a$ or $a^*$, to be the sum of the charge of the $\ell$ first terms: 
\begin{equation}\label{qell}
 q_\ell= \sum_{s=1}^\ell q(c_{i_s}).
\end{equation}
 Then, for a sequence of operators describing a $j$-block followed by a $i$-block (with $j>i$), the partial charge $q_\ell$ must satisfy $0\leq q_\ell \leq j$ for all values of $\ell$.
This is the basic requirement to be imposed in block mixing to ensure the block identities \cite{JMop}. Again, identical configurations are not counted twice. 

 The insertions of the $i$-block within the substring ${a^*}^j$ lead to configurations of the form:
\begin{equation}\label{inse}
{a^*}^{n}    ( {a^*}^i\, a^i)\,     {a^*}^{j-n}\, a^j  \qquad (0\leq n\leq j),
\end{equation}
(where the inserted block is delimitated by parentheses). The largest partial charge $q_{2j-n+i}$ is $i+n$ and in order to respect the bound $q_{2j-n+i} \leq j$, we need to have $i+n\leq j$. 
This excludes the insertion at the center of the $j$-block ($n=j$, which would correspond to a single block of type $i+j$) and its near  vicinity ($j-i<n<j$). The deepest penetration of the $i$-block within the $a^*$-side  of the $j$-block corresponds to the configuration with $n=j-i$:
\begin{equation}\label{insea}
{a^*}^{j-i}   ({a^*}^i\,  a^i) \,   {a^*}^{i}\,      a^j\, 
\end{equation}
There are thus $j-i+1$ distinct insertions within the substring ${a^*}^j$, counting the one with $n=0$ (which is the original configuration where the $i$-block follows the $j$-block). The insertions of the $i$-block within the substring ${a}^j$ are of the form
\begin{equation}\label{syma}
{a^*}^{j} \,{a}^{j-n'}\,  ({a^*}^i\,    a^i)\, a^{j-n'}\,    \qquad (0\leq n'\leq j).
\end{equation}
The strongest  constraint on the partial charge is $q_{j-n'+i}=j-n'+i\leq j$, which forces $n'\geq i$. 
There are also $j-i+1$ possible insertions on this $a$ side.
 However,  the case $n'=i$:
 \begin{equation}\label{symb}
{a^*}^{j}\, {a}^{j-i}\,  ({a^*}^i\,a^i)\,   a^{j-i}\,  
\end{equation}
when dropping the unessential parentheses,   leads to a configuration  identical to one already considered, cf. (\ref{syma}), and it should not be counted twice. The total number of distinct cases is thus $2(j-i)+1$. 

The successive displacements of the $i$-block within the $j$-block are considered from the left to the right. Once the closest-to-the-center penetration configuration is reached, which is (\ref{syma}),  it is exchanged with its symmetrical version (\ref{symb}) before the next displacement is considered. The next move starts from there.  
Each unit displacement of the $i$-block within the $j$-block  increases the weight by 1. This is rather obvious in the operator formalism (cf. \cite{JMop}) but since this has already been established from the paths, it will not  be proved again. The combinatorial factor for $p_i$ $i$-blocks inserted in $p_j$ $j$-blocks, keeping track of the relative weight, is 
\begin{equation}
\begin{bmatrix} p_i + 2(j-i)p_j\\ p_i\end{bmatrix}.
\end{equation} 
For $i=1, j=2$, this is the announced result (\ref{comb}).

Note that once the generating factor is obtained for a given block content, including those factors  that account for the block mixing, we need to sum over all possible values of the $p_j$, with $1\leq j\leq k$. This justifies preserving the block identities in the mixing process.

\subsection{From block mixing to particle penetration}

As previously claimed, the successive block insertions match  the successive steps of particle penetration, with the shape of the smaller particle alternating between its  deformed and undeformed version. 
%This claimed has been justified previously from the compatibility of the combinatorial analysis but it is most naturally seen to be a consequence of the block  mixing. 
This is most neatly illustrated by means of an example.
For instance the block representation of the various paths illustrating the penetration of a particle 3/2 within a particle 5/2 are as follows:
\begin{equation}
 (a^*a)a^*a^*a^*aaa,\quad 
 a^*(a^*a)a^*a^*aaaa,\quad 
a^*a^*(a^*a)a^*aaa,
\quad a^*a^*a^*aa(a^*a)a,
\quad a^* a^* a^* aaa (a^*a).
\end{equation} 
These are in respective correspondence with the paths illustrated in Fig. \ref{fig5}d.
In particular, in the path version of the second and the fourth block configurations  above, the  3/2 particle has its top  H edge moved to its  front or its rear.

\subsection{Operator basis}

As in \cite{JMop}, it is not difficult to write the conditions defining an operator basis in terms of constraints to be  imposed on successive modes of operators acting on the vacuum state.  The conditions can be summarized as follows:
\begin{align}\label{basis}  c_{i_m}c_{i_{m-1}} \cdots c_{i_1} :&\quad \text{$c$ is either $a$ or $a^*$, with $q(a)=-q(a^*)=1$,}\nonumber 
\\
%&\quad \text{$b$ acts on a maximum and $b^*$ on a minimum},\nonumber \\
%& \quad i_{s+1}-i_s\geq n+ \frac12+\frac14(1-q(c_{i_{s+1}})\, q(c_{i_s})), 
%
&  \quad 0\leq q_\ell\leq k-1,\quad \text{with $q_1=1$ and $q_m=0$},\nonumber\\ 
& \quad \text{for $q_s$ even}:  \quad i_{s+1}-i_s = 1+2n+\frac12(1-q(c_{i_{s+1}})\, q(c_{i_s})),\nonumber \\
& \quad \text{for $q_s$ odd}:  \quad\;  i_{s+1}-i_s = n+\frac12(1-q(c_{i_{s+1}})\, q(c_{i_s})),
\end{align}
where $n$ is any non-negative integer and $q_\ell$ is defied in (\ref{qell}).
The vacuum character is the generating function of all the sequences of operators subject to (\ref{basis}).

\section{Conclusion}

We have presented a fermi-gas description of the RSOS path representation of the superconformal unitary minimal models along the lines of  {\cite{OleJS}.  The finitized fermionic characters obtained here are new. However, their infinite length limit reproduce the formulae  previously found in \cite{BG,Sch}.  Since our interest lies essentially in the latter and that our method does not produce novel expressions for the conformal characters, we have 
restricted our presentation to paths with the simplest boundary conditions, those pertaining to the vacuum  module.\footnote{See however App. A for an interesting consequence of the novelty of these  finite characters in the context of parafermionic conformal theories.}

 The main interest of the present work is methodological: it shows that the fermi-gas technique  -- which is  fully constructive --  can be extended to models other than the Virasoro unitary ones.  The resulting positive multiple-sum character is thus formulated in terms of data that have a clear particle meaning in the path context.
 
% {\bf Possible addition: comments on the finite version vs the parafermionic case.}

 The generalization of the fermi-gas method from the Virasoro to the super-Virasoro case is not quite straightforward, however, since the particle interpretation is not completely  obvious in presence of horizontal edges within the path. The key hint in that regard comes from the (non-local)  operator representation of the path displayed in section 4, which generalizes our recent work \cite{JMop}. Retrospectively, this application is the best concrete motivation for  introducing these operators.

The unitary Virasoro and super-Virasoro minimal models are the first two diagonal cosets in the series $\widehat{su}(2)_k\oplus \widehat{su}(2)_n/ \widehat{su}(2)_{k+n}$, corresponding to the cases $n=1,2$ respectively. The character of all these series are combinatorially described by the RSOS configuration sums of the general class of models solved in \cite{Kyoto}.
A natural line of generalization is to extend the present fermi-gas analysis to the cases $n>2$. 

On the more combinatorial side, the taming of paths with H edges (which are actually restricted Motzkin paths with  a special weight  function) renders  tractable  the  study of generalized Bressoud  paths \cite{BreL}, where the H edges, originally confined to the horizontal axis, would be allowed at any height.  
But what is probably more interesting, is that, when combined with a simple path bijection, paths with H edges can be turned into a powerful tool for a fresh combinatorial analysis of the standard Forrester-Baxter paths. We will report elsewhere on this issue.

\appendix

\section{From the finitized RSOS characters to the parafermionic characters}

%\subsection{The $\z_{k+2}$ parafermionic characters}

The finitized characters in regime III are related to those of regime II by the duality transformation $q\rw q^{-1}$ \cite{Sch} (see also \cite{OleJS, FWa,JSTAT}). The conformal characters obtained by duality correspond  to the limit $L\rw \y$ but now evaluated with $n_{k/2+1}\rw \y$. The dual characters are those of the $\z_{k+2}$ parafermionic theories.  However, before this limit is taken, a correcting $L$-dependent factor needs to be introduced. Actually, the transformed character must be multiplied by  $q^{kL^2/2(k+2)}$. 
With these transformations, we  then  recover the character of the parafermionic vacuum  module, denoted $\chi_0^{(k+2)}$:
\begin{equation}
 \chi_0^{(k+2)}(q)= \lim_{L\rw \y} q^{\frac{k}{2(k+2)}L^2}\, {\tilde\chi}_{1,1}^{(k+2,k+4)}(q^{-1};L).
 \end{equation} 
The resulting expression  is (with $n_{i/2}$ replaced by $m_i$):
%in a somewhat modified form, namely a
\begin{equation}
%\lim_{L\rw \y} q^{\frac{(k-2)}{2k}L^2}\, {\tilde\chi}_{1,1}^{(k+2,k+4)}(q^{-1};L)= 
\chi_0^{(k+2)}(q)= \sum_{m_1,\cdots ,m_{k+1}=0}^\y \frac{q^{N_1^2+N_2^2+\cdots +N_{k+1}^2-\frac1{k+2}N^2+\frac12(m_1-m_1^2)}} {(q)_{m_1} (q^2;q^2)_{m_2}(q)_{m_3}\cdots (q)_{m_{k+1}} }
\end{equation} 
with 
\begin{equation}
N_i = m_i+\cdots +m_{k+1}\qquad\text{and} \qquad  N= \sum_{i=1}^{k+1} im_i.\end{equation}
This  is to be compared with the standard form \cite{LP} (see also \cite{JMqp}):
\begin{equation}
\chi_0^{(k+2)}(q)=\sum_{m_1,\cdots ,m_{k+1}=0}^\y \frac{q^{N_1^2+N_2^2+\cdots + N_{k+1}^2-\frac1{k+2}N^2}} {(q)_{m_1} (q)_{m_2}(q)_{m_3}\cdots (q)_{m_{k+1}} }
.\end{equation} 
The equivalence of these  two forms boils down to 
the identity
\begin{equation}\label{ide}
\sum_{m_1,m_{2}=0}^\y \frac{q^{\Lambda+ \frac12(m_1-m_1^2)}} {(q)_{m_1} (q^2;q^2)_{m_2} } = \sum_{m_1,m_{2}=0}^\y
 \frac{q^{\Lambda}} {(q)_{m_1} (q)_{m_2} } ,
 \end{equation}
where
\begin{equation}
\Lambda = \frac{1}{k+2}\,\l((k+1) m_1^2+ {2k}(m_1 m_2 + m_2^2)\r)+(m_1+2m_2)\,\Delta,
\end{equation}
for any $\Delta$.

The direct verification of this identity would be another validation of our expression for the finite characters.  Warnaar \cite{OW} offers us a simple but clever proof of (\ref{ide}). The first step amounts to transform the identity into a form that makes transparent its independence upon both $k$ and $\Delta$. For this, we
replace $q^{-1/(k+2)}$ by $z$ and $q^{\Delta}$ by $y$ and get:
\begin{equation}
\sum_{m_1,m_2=0}^\y \frac{ q^{(m_1+m_2)^2+m_2^2+\frac12(m_1-m_1^2)}\,
z^{(m_1+2m_2)^2}\, y^{m_1+2m_2} } { (q)_{m_1} (q^2;q^2)_{m_2} }
=
\sum_{m_1,m_2=0}^\y  \frac{ q^{(m_1+m_2)^2+m_2^2}\,
z^{(m_1+2m_2)^2} \, y^{m_1+2m_2}  } {  (q)_{m_1} (q)_{m_2} }.
\end{equation}
This can be rewritten as 
\begin{equation}
\sum_{\ell=0}^\y z^{\ell^2} \, y^\ell \sum_{\substack{m_1,m_2\geq 0\\ m_1+2m_2=\ell}} \frac{q^{(m_1+m_2)^2+m_2^2 +\frac12(m_1-m_1^2) } }
{ (q)_{m_1} (q^2;q^2)_{m_2} }
=\sum_{\ell=0}^\y z^{\ell^2 } \, y^\ell
\sum_{\substack{m_1,m_2\geq 0\\ m_1+2m_2=\ell}} \frac{q^{(m_1+m_2)^2 +m_2^2 } } { (q)_{m_1} (q)_{m_2} },
\end{equation} 
which reduces to
\begin{equation}
\sum_{\substack{m_1,m_2\geq 0\\ m_1+2m_2=\ell}} \frac{q^{(m_1+m_2)^2+m_2^2 +\frac12(m_1-m_1^2)} }
{ (q)_{m_1} (q^2;q^2)_{m_2} }
=
\sum_{\substack{m_1,m_2\geq 0\\ m_1+2m_2=\ell}} \frac{q^{(m_1+m_2)^2 +m_2^2 } } { (q)_{m_1} (q)_{m_2} }.
\end{equation}
Eliminating $m_1$ and renaming $m_2$ as $m$ give the following finite-sum identity:
\begin{equation}\label{bas}
\sum_{m=0}^{\ell/2} \frac{ q^{\frac12\ell(\ell+1)-m} }{ (q)_{\ell-2m} (q^2;q^2)_{m} }
=
\sum_{m=0}^{\ell/2}\frac{ q^{\ell^2-2\ell m+2m^2}} { (q)_{\ell-2m} (q)_{m} }.
\end{equation}
%This identity is ok : I have checked it extensively. It is useful to rewrite it with $\ell$ replaced by $2\ell$: it becomes
%\begin{equation} \label{iden1}
%\sum_{m=0}^{\ell} \frac{ q^{-m} }{ (q)_{2\ell-2m} (q^2;q^2)_{m} }=
%\sum_{m=0}^{\ell}\frac{ q^{2\ell^2-\ell-2\ell m+2m^2}} { (q)_{2\ell-2m} (q)_{m} }
% \end{equation}
For the demonstration of this identity, we will proceed in two steps, by considering both parities of $\ell$ separately.

Take first the case where $\ell$ is even. Set $\ell= 2k$, multiply both sides by $q^{-2k^2}$  and
then replace $m\to k-m$. This yields:
\begin{equation} 
 \sum_{m=0}^k \frac{q^m}{(q)_{2m}(q^2;q^2)_{k-m}}=\sum_{m=0}^k \frac{q^{2m^2}} {(q)_{2m}(q)_{k-m}} .
\end{equation}
At his point, the strategy of the proof is to try to reinterpret this as an identity for  suitably specialized  basic hypergeometric series. With this goal in mind, we multiply both sides by $z^k$  and sum $k$ over all integers $\geq 0$.
We then interchange the summations order and use the fact that $1/(q)_n=0$ if $n<0$  to reset the starting point of the sum over $k$  to $m$. 
The two $k$-summations can then be performed explicitly using 
Euler's $
q$-exponential sum (cf. \cite{Andr} eq. (2.2.5) or \cite{GR} eq. (II.1)):
\begin{equation} \label{euler}
\sum_{m=0}^\y \frac{t^n}{(q)_n}= \frac1{(t)_\y}.
\end{equation}
 The result is
\begin{equation} 
\frac1{ (z;q^2)_{\infty}}\sum_{m=0}^\y \frac{(zq)^m} { (q)_{2m}} =\frac1{(z)_{\infty} }
\sum_{m=0}^\y \frac{z^m q^{2m^2}}{(q)_{2m}},
\end{equation}
which we can rewrite under the form:
\begin{equation} \label{iden}
\sum_{m=0}^\y \frac{(zq)^m} { (q)_{2m}} =\frac{ (z;q^2)_{\infty}}{(z)_{\infty} }
\sum_{m=0}^\y \frac{z^m q^{2m^2}}{(q)_{2m}}= \frac{ 1}{(zq;q^2)_{\infty} }
\sum_{m=0}^\y \frac{z^m q^{2m^2}}{(q)_{2m}}.
\end{equation}
We next rewrite the left-hand-side in terms of 
%But this follows from a special Heine transformation 
of the $_2\phi_1(a,b;c;q;z)$ series defined as \cite{GR}:
\begin{equation} 
_2\phi_1(a,b;c;q;z)= \sum_{n=0}^\y\frac{(a)_n\, (b)_n \, z^n}{(c)_n \, (q)_n}\equiv  \sum_{n=0}^\y\frac{(a;q)_n\, (b;q)_n \, z^n}{(c;q)_n \, (q;q)_n}.
\end{equation}
Using the simple identity:
$ (q)_{2m}= (q;q^2)_m\, (q^2;q^2)_m$, we see that 
\begin{equation} 
\sum_{m=0}^\y \frac{(zq)^m} {(q;q^2)_m(q^2;q^2)_m} = {_2\phi_1}(0,0;q;q^2;zq).
\end{equation}
The identity (\ref{iden}) follows from a special Heine transformation, namely (see
\cite{GR} eq. (III.3)):
\begin{equation}\label{heine}
_2\phi_1(a,b;c;q;z)= \frac{(abz/c;q)_\y}{(z;q)_\y}    
\, {_2\phi_1}(c/a,c/b;c;q;abz/c).
\end{equation}
In the present context, we must set $a,b\to 0 $ through  a limiting process,
%$a\rw 0, \, b\rw 0$
using
\begin{equation}\lim_{a\rw 0}\,  (q/a)_n\, a^n = (-1)^n q^{n^2}.
\end{equation}
Hence, the right-hand-side of (\ref{heine}), with $a,b\to 0, c=q,z\to zq$ and $q\to q^2$, yields directly  the second member of (\ref{iden}). This completes the proof of the identity (\ref{bas}) when $\ell$ is even.

Consider next the case where $\ell= 2k+1$.
Once this substitution is done in (\ref{bas}), we multiply the  two sides by $q^{-2k^2-2k}$ and  let again $m\to k-m$
 to obtain:
\begin{equation}\sum_{m=0}^k\frac{ q^m }{(q)_{2m+1}(q^2;q^2)_{k-m}} =
\sum_{m=0}^k\frac{q^{2m(m+1)}} {(q)_{2m+1}(q)_{k-m}Ê}.
\end{equation}
We then proceed as before: multiply both sides by $z^k$, sum over $k\geq 0$, and use (\ref{euler}). That gives
\begin{equation}
\sum_{m=0}^\y \frac{(zq)^m }{(q)_{2m+1} } = \frac{1}{(zq;q^2)_{\infty} }
\sum_{m=0}^\y \frac{(zq^2)^m q^{2m^2}Ê}{(q)_{2m+1} }.
\end{equation}
Next, we use another simple identity: $(q)_{2m+1}= (1-q)\, (q^2;q^2)_m \, (q^3;q^2)_m$, and cancel the factor $(1-q)$ from both sides to get
\begin{equation}
\sum_{m=0}^\y \frac{(zq)^m }{(q^2;q^2)_m (q^3;q^2)_m} =\frac{1}{(zq;q^2)_{\infty} }
\sum_{m=0}^\y \frac {(zq^2)^m q^{2m^2}}{(q^2;q^2)_m (q^3;q^2)_m} .
\end{equation}
But this is again a special case of Heine's transformation (\ref{heine}):
\begin{equation}\label{heinea}
_2\phi_1(0,0;q^3;q^2;zq)= \frac{1}{(zq;q^2)_\y} \, \lim_{a\rw 0}   
\, {_2\phi_1}(q^3/a,q^3/a;q^3;q^2;a^2 z/q^2).
\end{equation}
This completes the proof of (\ref{bas}), hence of (\ref{ide}).

%\n $\bullet$ Ajouts possibles: autres modules ET la descripion complete de la base
%YYY

\vskip0.3cm
\noindent {\bf ACKNOWLEDGMENTS}

We thank O. Warnaar for generously providing us with the proof of (\ref{ide}) that has been presented in App. A and for clarifications concerning the identity (\ref{Siden}).
% and very useful comments.
% - and thus for the content of App. A  starting from (\ref{ide}) to the end -- and very useful comments. 
This  work is supported  by NSERC.

\end{document}